\documentclass{iopart}[12pt]
\usepackage{iopams}
\usepackage{epsf}
\eqnobysec
\def\r{\mathbb R}                  
                              
\def\d{\partial}
\def\nb{\nabla}
\def\fr{\frac}
\def\be{\begin{equation}}
\def\ee{\end{equation}}
\def\bea{\begin{eqnarray}}
\def\eea{\end{eqnarray}}
\def\bnr{\begin{eqnarray*}}
\def\enr{\end{eqnarray*}}
\def\N{\hfill \raisebox{1mm}{\framebox{\rule{0mm}{1mm}}}}
\def\T{{\bf T}}
\def\Su{{\bf S}}
\def\p{{\bf P}}

\def\G{{\bf g}}

\def\rmg{{\rm g}}

\def\g{\gamma}

\def\z{\btheta}
\def\s{\sigma}
\def\O{{\bf\Omega}}
\def\Chr{\Gamma}
\def\Sg{{\bf\Sigma}} 
\def\f{\varphi}

\def\P{{\bf Proof :} \hspace{3mm}}
\def\k{\vec{\bf k}}

\def\Gl{{\cal G}}
\def\l{\lambda}
\def\xiv{\vec{\bxi}}
\def\lie{{\pounds}_{\xiv}\, }
\def\liea{{\pounds}_{\xiv_1}\, }
\def\lieb{{\pounds}_{\xiv_2}\, }
\def\lieu{{\pounds}_{\u}\, }

\def\vu{\bm{u}}
\def\e{{\bf e}}

\def\u{\vec{\bm{u}}}

\def\K{{\bm k}}

\def\vecX{{\mathfrak X}}
\def\det{\mbox{det}}

\newcommand{\bm}[1]{\mbox{\boldmath $#1$}}
\newtheorem{defi}{Definition}[section]
\newtheorem{theo}{Theorem}[section]
\newtheorem{coro}{Corollary}[section]
\newtheorem{prop}{Proposition}[section]
\newtheorem{lem}{Lemma}[section]

\begin{document}
\title[Bi-conformal vector fields]
{Bi-conformal vector fields and their applications}

\author{Alfonso Garc\'{\i}a-Parrado and Jos\'{e} M M Senovilla}

\address{F\'{\i}sica Te\'{o}rica,
 Universidad del Pa\'{\i}s Vasco, Apartado 644, 48080 Bilbao, Spain}

\eads{\mailto{wtbgagoa@lg.ehu.es} and \mailto{wtpmasej@lg.ehu.es}}

\begin{abstract}
We introduce the concept of {\em bi-conformal transformation}, as a 
generalization of conformal ones, by allowing two orthogonal parts of 
a manifold with metric $\G$ to be scaled by different conformal factors. In 
particular, we study their infinitesimal version, called
bi-conformal vector fields. We show that these are characterized by the differential conditions 
$\lie \p\propto \p$ and $\lie {\mathbf \Pi} \propto {\mathbf \Pi}$ where 
$\p$ and ${\mathbf \Pi}$ are
orthogonal projectors ($\p +{\mathbf \Pi}=$ \G). 
Keeping $\p$ and ${\mathbf \Pi}$ fixed, the set of bi-conformal vector fields is a Lie 
algebra which can be finite or infinite dimensional according 
to the dimensionality of the projectors. We determine (i) when an infinite-dimensional case 
is feasible and its properties, and (ii) a {\em normal system} 
for the generators in the finite-dimensional case. Its integrability conditions are also analyzed, 
which in particular provides the maximum number of linearly independent 
solutions. We identify the corresponding maximal spaces, and show
a necessary geometric condition for a metric tensor to be a double-twisted 
product. More general ``breakable'' spaces are briefly considered.
Many known symmetries are included, such as conformal Killing 
vectors, Kerr-Schild vector fields, kinematic self-similarity, causal symmetries, 
and rigid motions.
\end{abstract}

\section{Introduction}
Symmetry transformations have been a subject of research 
over the years. In General Relativity they have been used 
for different purposes, ranging from the classification of exact solutions of the field equations to the 
generation techniques for new solutions \cite{KRAMERS}. In this work, we are interested in 
the study of continuous transformation groups with certain properties acting 
on a metric manifold (see \cite{HALL1,HALL2} for
a precise definition of this). A classification 
of the outstanding cases can be found in \cite{LEVINE} where they are sorted according to the 
{\it differential conditions} complied by the infinitesimal generators. 
This condition involves the Lie derivative of the metric tensor or other 
geometric objects ---such as the connection or the 
curvature tensor---. The symmetries classified in \cite{LEVINE} have received 
a great deal of attention. However, as a matter of fact, it is difficult to find in the literature studies of 
symmetries characterized by other differential conditions.
Some examples can be found in 
\cite{KERR-SCHILD,sergi,TSAMPARLIS,ZAFIRIS,LETTER}. 

In this paper we will pursue this line of research 
and  present a new type of group of transformations: those 
diffeomorphisms which scale two pieces of the metric tensor by 
unequal factors. We call them {\em bi-conformal transformations}. 
We will not restrict our presentation to four-dimensional spacetimes, 
so that our results will be valid in any $n$-dimensional differential manifold $V$ 
with a smooth metric tensor $\G$ of any signature. 
Bi-conformal transformations can be univocally characterized by a {\em symmetric square root} of 
$\G$ (see next section for its definition; this is a Lorentz tensor in Lorentzian signature, see 
\cite{PI,PIII}), or equivalently by two complementary orthogonal 
projectors. The infinitesimal generators are well defined and we
present the necessary and sufficient differential conditions they 
fulfill, which involve the Lie derivatives of the metric tensor and of the 
square root, or equivalently of the two projectors. 
This differential condition can be understood as stating that the
generating vectors, called {\em bi-conformal vector fields}, 
are generalized conformal motions for both projectors. 
The properties of bi-conformal vector fields are studied, and we show 
that they constitute a Lie algebra which can be finite or infinite 
dimensional. We identify the cases where the latter case may happen. 
We also prove that, in the former case, a normal system---for 
$p,n-p\neq 1,2$ where $p$ is the trace of one of the projectors---can be 
achieved, so that the integrability conditions and the maximum number 
of linearly independent bi-conformal vector fields are found. This 
turns out to be $(p+1)(p+2)/2+(n-p+1)(n-p+2)/2$. We show that this maximum number is 
attained in double-twisted product spaces with flat leaves (i.e. a 
metric breakable in two conformally flat pieces where the conformal factor of 
each part depend on all the coordinates of the manifold).
We also find a necessary geometric condition for a space to admit 
such form in local coordinates.        

The outline of the paper is as follows: in section \ref{PURECAUSAL} we set the notation and 
study the basic properties of square roots. Bi-conformal vector fields are introduced in section \ref{BI} 
whereas in section \ref{bi-trans} we study groups of 
bi-conformal transformations and give the general form taken by the metric tensor in a 
coordinate system adapted to the symmetry. The Lie algebra 
of bi-conformal vector fields is the subject of section \ref{LIE}. 
The normal system for the finite-dimensional case
is obtained in section \ref{highest} together with the highest dimension of the Lie algebra. 
Part of the integrability conditions of the 
afore-mentioned equations are considered in section \ref{conditions}. Finally explicit examples of bi-conformal 
vector fields are presented in section \ref{examples}. 

\section{Preliminaries.}
\label{PURECAUSAL}
We start by setting the notation and conventions to be used in this work. $(V,\G)$ will stand for 
a smooth $n$-dimensional manifold with metric $\G$. The metric signature is arbitrary although in some 
of our results we will specialize $\G$ to a Lorentzian metric (signature convention 
$(+,-,\dots,-)$) in order to highlight the applications to $n$-dimensional spacetimes. 
Latin characters running from $1$ to $n$ will be used for tensor indexes.
Vectors and contravariant (covariant) tensors will be denoted with arrowed (un-arrowed) boldface 
characters whenever they are expressed in index-free notation. As usual contravariant and covariant tensors 
are related by the rule of raising and lowering of indexes. 
Thus if $\vec{\T}$ is a rank-$r$ contravariant tensor
we will use the same un-arrowed symbol $\T$ for the tensor obtained by lowering all the indexes. 
Index notation though will be used in most of the paper for tensors. 
Round and square brackets enclosing indexes 
will stand for symmetrization and anti-symmetrization respectively. We review next very briefly some basic 
geometric concepts in order to show the notation we use.
In $(V,\G)$ we define the tangent space $T_x(V)$ at a point $x\in V$,
the tangent and cotangent bundles $T(V)$, $T^*(V)$ and the bundle $T^r_s(V)$ of 
$r$-covariant $s$-contravariant tensors in the usual way. 
All differentiable sections of the bundle (vector fields) 
$T(V)$ and other tensor bundles will be assumed smooth. We will use the same 
 notation for sections as for vectors and tensors unless the context requires 
to use different notations. 

Recall that any $C^1$ vector field 
$\xiv$ on a differentiable manifold defines a local group of {\em local} 
diffeomorphisms $\{\f_s\}$ that is to say, each member $\f_s$ of the family is a local diffeomorphism and
$\f_0$ is the identity on $V$. In addition to this 
 $(\f_{s_1}\circ\f_{s_2})(x)=\f_{s_1+s_2}(x)$ holds whenever all the objects appearing in 
both sides of the equality make sense. In local coordinates $\{x^a\}$ we have the correspondence
$$
\xi^a(\f_s(x))=\fr{d\f^a_s(x)}{ds},\ \ \ \ \f_0(x)=x
$$  
which relates the vector field $\xiv$ with its generated local group of local diffeomorphisms 
by means of standard theorems on differential equations. When this local group is formed by global 
diffeomorphisms $\f_s:V\rightarrow V$, $s\in I$ then $\xiv$ is called a {\em complete} vector field
being $I$ an interval of the real line containing $0$.
For a Hausdorff manifold this is equivalent to $I=\r$ or in other words the integral curves of 
this vector field can be extended to every value of their parameter.
 As is well known vector fields can be regarded as 
differential operators 
acting on the set of smooth functions of the manifold $V$ and in this picture
 the Lie bracket of two $C^1$ vector fields $\xiv_1$, $\xiv_2$ is 
$$
[\xiv_1,\xiv_2](f)\equiv\xiv_1(\xiv_2(f))-\xiv_2(\xiv_1(f)).
$$
 The set of smooth vector fields generates an infinite dimensional 
Lie algebra by means of the Lie bracket operation. This is usually denoted by 
$\vecX(V)$ and not all the elements of $\vecX(V)$ are complete unless the 
dimension of the Lie algebra is finite. 
We can use the (local) one-parameter group of diffeomorphisms $\{\f_s\}$ generated by $\xiv$ to 
define, for any $\vec{\T}\in T^r(V)$, the family of tensor fields $\f'_s\vec{\T}$ where $\f'_s$ 
is the push-forward of $\f_s$.
There is an obvious counterpart for covariant tensor fields $\T$ using the pull-back $\f^*_s$.
The Lie derivative of $\vec{\T}$ $(\T)$ is another tensor field of the same rank defined by
$$
\lie\vec{\T}=\lim_{s\rightarrow 0}\fr{\f'_{-s}\vec{\T}-\vec{\T}}{s},\ \ 
\lie\T=\lim_{s\rightarrow 0}\fr{\f^*_s\T-\T}{s}.
$$
All these geometric definitions are well known and can be found in many text books 
(see e. g. \cite{CHERN}).  
\subsection{Square roots.} 
We present next a concept which will play a very important role in this work.
\begin{defi}
Let $\G|_x$ be a nondegenerate bilinear form defined on $T_x(V)$. A symmetric tensor $S_{ab}$ on $T_x(V)$
 is called a square root of $\G|_x$ if 
$$
S_{ap}S^p_{\ b}=\rmg_{ab}.
$$
\label{square-root}
\end{defi}
If $\rmg_{ab}$ has Lorentzian signature then the tensor $S_{ab}$ is called a {\em Lorentz tensor}. 
Clearly Lorentz tensors with an index raised are involutory Lorentz transformations 
(a linear transformation is said to be involutory if its inverse is the transformation itself) but
they can also be characterized as {\it superenergy tensors} of certain normalized simple forms 
as it was shown in \cite{PI}:
\begin{prop}
Every Lorentz tensor $S_{ab}$ is proportional to the superenergy tensor of a simple form $\O$. 
\label{converso}
\end{prop}
Without going into further details which are beyond the scope of 
this paper, we will just write down the definition of the superenergy tensor $T_{ab}\{\O\}$ 
of a $p$-form $\O$ \cite{SUP,PI}
\be
 T_{ab}\{\O\}=\fr{(-1)^{p-1}}{(p-1)!}
\left(\Omega_{ac_2\dots c_p}\Omega_{b}^{\ c_2\dots c_p}-
\fr{1}{2p}\rmg_{ab}\Omega_{c_1\dots c_p}\Omega^{c_1\dots c_p}\right).
\label{superenergy}
\ee 
The form $\O$ is said to be simple if it can be decomposed as the wedge product 
of 1-forms and 
the normalization required is
\be
 \O\cdot\O\equiv \Omega_{c_1\dots c_p}\Omega^{c_1\dots 
 c_p}=2p!(-1)^{p-1}.\label{norma}
\ee
 Denoting by $\K_1,\dots,\K_p$ 
a set of 1-forms such that $\O=\K_1\wedge\dots\wedge\K_p$ we have that $Span\{\k_1,\dots,\k_p\}$ and 
$\perp Span\{\k_1\dots,\k_p\}$ are 
the only eigenspaces of $S^a_{\ b}$ with corresponding eigenvalues $+1$ and $-1$ respectively. 
Another important property is the invariance $T_{ab}\{\O\}=T_{ab}\{\pm*\O\}$ where $*\O$ is the 
hodge dual of $\O$. 
Thus, whenever we speak of the Lorentz tensor of a $p$-form this must be understood 
up to duality and sign. With the normalization chosen above, $Span\{\k_1,\dots,\k_p\}$ is a timelike
subspace ($\O$ is ``timelike'') and $\perp Span\{\k_1,\dots,\k_p\}$ is its spacelike complement 
($*\O$ is spacelike). Some of these results can be translated to arbitrary square roots.
\begin{prop}
Every square root $S^a_{\ b}$, regarded as
an endomorphism, has $+1$ and $-1$ as unique eigenvalues. Furthermore 
$S^a_{\ b}$ is diagonalizable so the space $T_x(V)$ decomposes in 
a direct sum of the two eigenspaces associated to the eigenvalues $+1$ and $-1$.   
\label{spectral}
\end{prop}
\P The first statement of the proof is trivial. To prove the second one
we must note that the endomorphisms $P^{a}_{\ b}=(\delta^a_{\ b}+S^{a}_{\ b})/2$, 
$\Pi^{a}_{\ b}=(\delta^a_{\ b}-S^{a}_{\ b})/2$
are idempotent, their composition in any order gives the zero endomorphism, 
both have vanishing determinant and $\delta^{a}_{\ b}=P^{a}_{\ b}+\Pi^{a}_{\ b}$.
From this we deduce that the direct sum of the eigenspaces of $S^a_{\ 
b}$, with eigenvalues $+1$ and $-1$,
is the total space or in other words $S^a_{\ b}$ is diagonalizable.\N

The following straightforward corollary is deduced from this proposition. 
\begin{coro}
The tensors
\be
P_{ab}=(\rmg_{ab}+S_{ab})/2,\ \ \Pi_{ab}=(\rmg_{ab}-S_{ab})/2
\label{losproyectores}
\ee
constructed from any square root $S_{ab}$ are orthogonal projectors on the eigenspaces of eigenvalue 
$+1$ and $-1$ respectively.
\end{coro}
For the case of Lorentzian signature the subspaces onto which $P_{ab}$ and $\Pi_{ab}$ project 
coincide with $Span\{\mu(\Su)\}$ and $\perp Span\{\mu(\Su)\}$ where $\mu(\Su)$ is the set of null vectors 
$\k$ such that $\Su(\k,\k)=0$ (see proposition A.3 of \cite{PIII} for further details). 

By means of previous results, we can now generalize proposition \ref{converso} to cases in which 
the metric tensor is not necessarily Lorentzian.
\begin{theo}
Any square root $S_{ab}$ can be written, up to sign, as in (\ref{superenergy}) 
for a simple form $\O$ normalized, up to sign, as in (\ref{norma}).
\label{S=T}
\end{theo}
\P Under the assumptions of this theorem a not very long calculation shows that for a $p$-form 
$\Omega_{a_1\dots a_p}$ we have
$$
T_{ap}\{\O\}T^p_{\ b}\{\O\}=\fr{(\O\cdot\O)^2}{(2p!)^2},
$$
(the procedure is the same as for the Lorentzian case, see the proof of proposition 3.2 in \cite{PI}).
Thus formula (\ref{superenergy}) still defines a square root if $\O\cdot\O=\pm 2p!$ in the 
case of a non-Lorentzian metric.  
Conversely if $S^a_{\ b}$  is a square root then proposition \ref{spectral} tells us that we can write 
$T_x(V)$  as a direct sum of the eigenspaces of $S^a_{\ b}$ with eigenvalues +1 and -1 denoted by 
$V_{+}$ and $V_{-}$ respectively. 
Furthermore $S^a_{\ b}$ is a symmetric endomorphism with respect to the metric $\G|_x$ so 
$V_{+}^{\perp}=V_{-}$. Now choose a basis $\{\k_1,\dots,\k_p\}$ of $V_{+}$ (we take it of dimension $p$)
 and calculate the
tensor $\T\{\Sg\}$ with $\Sg=\K_1\wedge\dots\wedge\K_p$. Use of formula (\ref{superenergy}) implies 
that $\k_j$ $j=1,\dots p$, are eigenvectors of $T^a_{\ b}\{\Sg\}$ with eigenvalue $(-1)^{p-1}\Sg\cdot\Sg/(2p!)$
and the same happens for any vector in $V_{-}$ (but now the eigenvalue has the opposite sign). 
The identity
$$
\fr{1}{(p-1)!}\Sigma^{ab_1\dots b_{p-1}}\Sigma_{bb_1\dots b_{p-1}}+\fr{|\det(\G)|}{\det(\G)(n-p-1)!}
(*\Sigma)^{ab_{p+2}\dots b_n}(*\Sigma)_{bb_{p+2}\dots b_n}=
\fr{\Sg\cdot\Sg}{p!}\delta^a_{\ b},
$$ 
must be used along the way to prove this. 
Hence the tensor $T^a_{\ b}\{\Sg\}$ has the same spectral properties as 
$S^a_{\ b}$ which is only possible if both are proportional. The proportionality 
factor can be fixed to
$+1$ or $-1$ by choosing the normalization $\Sg\cdot\Sg=\pm 2p!$.\N

\noindent
{\bf Remark.} Note that the sign of $\Sg\cdot\Sg$ is fixed and
characteristic to the subspace $V_+$ 
(if the metric has Lorentzian signature this sign allows us to decide the causal character of the subspace
$V_+$). One can thus choose the sign in the normalization condition
 $\Sg\cdot\Sg=\pm 2p!$ so that $\T\{\Sg\}=\Su$.      

\section{Bi-conformal vector fields} 
\label{BI}
\begin{defi}
 $\xiv$ is said to be a {\em bi-conformal vector field} if it fulfills the differential conditions 
\be
\lie\G=\alpha\G+\beta\Su,\ \ \lie\Su=\alpha\Su+\beta\G,
\label{fundamental}
\ee
where $\Su$ is a symmetric square root of $\G$ and $\alpha$, $\beta$ smooth functions.
\label{bi-conformal}
\end{defi}
Following \cite{KERR-SCHILD} the functions $\alpha$ and $\beta$ will be called {\it gauges} of the symmetry 
and their true relevance will become clear later. Suffice it to say here
they play a role analogous to the factor $\psi$ appearing in the differential condition $\lie\G=2\psi\G$ 
satisfied by conformal motions \cite{YANO}. 

If the signature admits null vectors then there is a variant of previous definition known as generalized 
Kerr Schild vector fields where $S_{ab}$ is no longer a square root but it 
takes the form $S_{ab}=k_ak_b$ with $k^ak_a=0$ (this can be also characterized as 
$S_{ap}S^p_{\ b}=0$). The explicit differential condition is now
\be
\lie\G=\alpha\G+\beta\K\otimes\K,\ \ 
\lie\K=\g\K.
\label{Ku-Ku}
\ee
Generalized Kerr-Schild vector fields are a generalization of Kerr-Schild vector fields studied in 
\cite{KERR-SCHILD} given by (\ref{Ku-Ku}) with $\alpha=0$. 
Both bi-conformal vector fields and generalized Kerr-Schild vector fields 
can be written in a gauge-free way as we show in the next theorem.
\begin{theo}
A vector field $\xiv$ is either a bi-conformal vector field or a generalized Kerr-Schild vector 
field if and only if
\be
(\lie\G\times\lie\G)\wedge\lie\G\wedge\G=0,\ \ (\lie\lie\G)\wedge\lie\G\wedge\G=0.
\ee
\label{gauge-free}
\end{theo}    
\noindent 
\P The inner product $\times$ of two rank-2 tensors $T_{ab}$ and $M_{ab}$ 
is defined by $(T\times M)_{\ ab}=T_{ac}M^{c}_{\ b}$ and their wedge product $\wedge$ is the typical 
wedge product in 
$T^0_2(V)$ considered as a vector space, e.g. $(T\wedge M)_{abcd}=T_{ab}M_{cd}-T_{cd}M_{ab}$. 
It is clear that the statement of this theorem is equivalent to demanding the existence of functions
$\l_1$, $\l_2$, $\l_3$ and $\l_4$ with the properties
\be
\lie\G\times\lie\G=\l_1\lie\G+\l_2\G,\ \ \lie\lie\G=\l_3\G+\l_4\lie\G,
\label{expanded-form}
\ee
so we will prove the equivalence of these equations to those characterizing bi-conformal vector
 fields and generalized Kerr-Schild 
vector fields. Straightforward calculations show that the above expressions are fulfilled with
$$
\l_1=2\alpha,\ \l_2=\beta^2-\alpha^2,\ \l_3=-\alpha^2+\beta^2+
\lie\alpha-\fr{\alpha}{\beta}\lie\beta,\ \l_4=\fr{1}{\beta}\lie\beta+2\alpha,
$$
if $\xiv$ is a bi-conformal vector field, and with
$$
\l_1=2\alpha,\ \l_2=-\alpha^2,\ 
\l_3=\lie\alpha-\fr{\alpha}{\beta}\lie\beta-2\g\alpha,\ 
\l_4=\fr{1}{\beta}\lie\beta+\alpha+2\g,
$$
for a generalized Kerr-Schild vector field. Conversely, assuming that (\ref{expanded-form}) holds 
and setting $\T=\lie\G-\fr{1}{2}\l_1\G$, the first of these equations implies that 
$\T\times\T\propto\G$ so that $\T$ is proportional to a square root of $\G$ or 
$\T\times\T=0$. In the former case, $\T=\beta\Su$ for some square root $\Su$
 whence $\lie\G=\alpha\G+\beta\Su$ with $\alpha=\fr{1}{2}\l_1$ whose 
substitution in the second equation of (\ref{expanded-form}) allows us to get
$$
\lie\Su=\mu_1\G+\mu_2\Su,\ \ \mu_1=\fr{1}{\beta}(\l_3+\l_4\alpha-\lie\alpha-\alpha^2),
\ \ \ \mu_2=\l_4-\fr{1}{\beta}\lie\beta-\alpha.
$$
The functions $\mu_1$ and $\mu_2$ can be further constrained by applying $\lie$ to the 
relation $\delta^a_{\ b}=S^a_{\ c}S^c_{\ b}$ getting
$\mu_1=\beta$, $\mu_2=\alpha$ from what we deduce that $\xiv$ is a bi-conformal 
vector field. In the case $\T\times\T=0$ then either $\T=0$, so that
$\xiv$ is a conformal Killing vector, or
$\T=\K\otimes\K$, with $\K$ null (this can only happen if the metric admits 
null vectors) so that
$$
\lie\K\otimes\K+\K\otimes\lie\K=\mu_1\G+\mu_2\K\otimes\K,
$$
which only makes sense if $\mu_1=0$. 
This gives the differential condition characterizing generalized Kerr-Schild vector fields  
at once.\N  
                                                                   
Bi-conformal vector fields will be studied paying special attention to their geometrical meaning.
To start with we show how the second equation of (\ref{fundamental}) can be rewritten in 
terms of the $p$-form giving rise to the square root $\Su$ of $\G$
 entering in the definition of bi-conformal vector fields.
\begin{prop}
The second equation of (\ref{fundamental}) admits the following equivalent forms 
\be
\lie\O=\fr{p}{2}(\alpha+\beta)\O,\ \ \ \lie\vec{\O}=-\fr{p}{2}(\alpha+\beta)\vec{\O},
\label{diff-form}
\ee
where $\O$ is a simple $p$-form such that $\Su=\T\{\O\}$ and the first of (\ref{fundamental}) is assumed.
\label{second-form}
\end{prop}
\P From equations (\ref{fundamental}) it follows, by means of $\lie\rmg^{ab}=-\alpha\rmg^{ab}-\beta S^{ab}$, that 
$\lie S^a_{\ b}=0$ so that 
for any vector $\k$ intervening as a factor in the $p$-vector $\vec{\O}$ we have $S^a_{\ p}\lie k^p=\lie k^a$ 
as follows by Lie 
derivating the equation $S^a_{\ p}k^p=k^a$. This means that $\lie k^a$ is also an eigenvector of $S^a_{\ b}$ and 
thus $\lie k^a$ is a 
linear combination of the vectors which build up $\vec{\O}$ whence $\lie\vec{\O}\propto\vec{\O}$. Using the
 relation 
between $\O$ and $\vec{\O}$ and the first of (\ref{bi-conformal}) 
we get a similar formula for $\lie\O$. The proportionality factors are 
fixed by Lie derivating the normalization condition on $\O\cdot\O$ written before, arriving thus at 
(\ref{diff-form}). Conversely, if any of (\ref{diff-form}) together with the first equation of 
(\ref{fundamental}) hold we can 
work out the other equation of (\ref{diff-form}) using the relation between $\O$ and $\vec{\O}$ together with 
$
\Omega_{a_1\dots a_p}S^{a_1}_{\ b_1}=\Omega_{b_1a_2\dots a_p}.
$
Then $\lie S_{ab}$ is calculated by just derivating (\ref{superenergy}) (as all the required Lie derivatives 
are known) getting the second equation of (\ref{fundamental}).\N  

\section{Bi-conformal transformations}
\label{bi-trans}
 The structure of the differential conditions for bi-conformal vector fields 
allows us to find explicit expressions for $\f^{*}_s\G$ and $\f^{*}_s\Su$.
To that end we rewrite equations (\ref{bi-conformal}) as 
\be
\lie(\G+\Su)=(\alpha+\beta)(\G+\Su),\ \ \ \lie(\G-\Su)=(\alpha-\beta)(\G-\Su).
\label{semi-conformal}
\ee
Observe that these are differential conditions on the projectors $P_{ab}$ 
and $\Pi_{ab}$, so that they are conformally invariant, and one 
obviously gets
\be
\lie P^a{}_{b} =\lie \Pi^a{}_{b}=0 .\label{mixed-inv}
\ee
Now if we take into account the identity
\be 
\fr{d(\f^*_s\T)}{ds}=\f^*_s(\lie\T),
\label{lie-property}
\ee
holding for every section $\T$ of $T^0_r(V)$, equations   
(\ref{semi-conformal}) can be integrated yielding
\bea
\f^{*}_s\G+\f^{*}_s\Su=
(\G+\Su)\exp\left[\int_0^sdt(\alpha(\f_t)+\beta(\f_t))\right]\label{finite-form1}\\
\f^{*}_s\G-\f^{*}_s\Su=(\G-\Su)\exp\left[\int_0^sdt(\alpha(\f_t)-\beta(\f_t))
\right],\label{finite-form2}
\eea
from what we get the formulae for $\f^{*}_s\G$ and $\f^{*}_s\Su$
\bea
\fl\f^{*}_s\G=\exp\left\{\int_0^sdt\ \alpha(\f_t)\right\}
\left[\cosh\left(\int^s_0dt\ \beta(\f_t)\right)\G+
\sinh\left(\int_0^sdt\ \beta(\f_t)\right)\Su\right],\\
\fl\f^{*}_s\Su=\exp\left\{\int_0^sdt\ \alpha(\f_t)\right\}
\left[\cosh\left(\int^s_0dt\ \beta(\f_t)\right)\Su+
\sinh\left(\int_0^sdt\ \beta(\f_t)\right)\G\right].
\eea
It is also possible to perform a similar derivation for generalized Kerr-Schild vector fields  
if the signature admits null vectors. The result is
\bnr
\fl\f^{*}_s\G=\exp\left[\int_0^sdt\ \alpha(\f_t)\right]\left(\G+\int_0^sdt\ 
\exp\left[-\int_0^tdt^{'}\alpha(\f_{t^{'}})\right]f^2(t)\beta(\f_t)
\K\otimes\K\right).\\
\fl\f^{*}_s\K=\K\exp\left[\int^{s}_{0}dt\ \g(\f_t)\right],\ \  
f(t)=\exp\left[\int^{t}_{0}du\ \g(\f_u)\right].
\enr

We thus see that bi-conformal vector fields can be characterized as 
{\em generalized} conformal motions\footnote{They are ``generalized'' 
in the sense that the conformal factors and gauges are functions on $V$, that is 
to say, they depend on {\em all} coordinates.} of both projectors $P_{ab}$ and 
$\Pi_{ab}$ of (\ref{losproyectores}) which states clearly the geometric interpretation of these vector fields. 
This interpretation will be 
further supported when we study the highest dimension of finite-dimensional Lie algebras of bi-conformal vector fields in section 
\ref{highest} (see below for the definition of Lie algebras of bi-conformal vector fields).  This characterization and equations 
(\ref{finite-form1}), (\ref{finite-form2})  lead us to the definition of {\em bi-conformal transformation}
\begin{defi}
 A diffeomorphism $\Phi:V \rightarrow V$ is called a bi-conformal transformation if there exists a pair of orthogonal
projectors $P_{ab}$ and $\Pi_{ab}$ with $\rmg_{ab}=P_{ab}+\Pi_{ab}$ such that 
$$
\Phi^*P_{ab}=\l_1P_{ab},\ \ \ \Phi^*\Pi_{ab}=\l_2\Pi_{ab},
$$ 
for some functions $\l_1$, $\l_2\in C^1(V)$. Equivalently, this can be rewritten in terms of the square root
$S_{ab}=P_{ab}-\Pi_{ab}$ as  
$$
\Phi^*\rmg_{ab}=\rho_1\rmg_{ab}+\rho_2S_{ab},\ \ \Phi^*S_{ab}=\rho_2\rmg_{ab}+\rho_1S_{ab},
$$
where $\rho_1=(\l_1+\l_2)/2$, $\rho_2=(\l_1-\l_2)/2$.
\label{bi-conformal-transformation}
\end{defi}

A clue for the interpretation of bi-conformal vector fields can be obtained by examining 
their effect on $\G$ in local coordinates adapted to $\xiv$.
\begin{prop}
Let $\xiv$ be a bi-conformal vector field and $\{x^1,x^i\}$, $i=2,\dots,n$ a 
local coordinate system adapted to $\xiv$ (i.e. $\xiv=\d/\d x^1$) 
around any non-fixed point of $\xiv$.
In these coordinates the metric takes the form
\be
\rmg_{ab}=e^AG^0_{ab}(x^i)+e^BG^1_{ab}(x^i), 
\label{coordinate-free1}
\ee 
where {\it (i)} $G^0_{ab}$ and $G^1_{ab}$ do not depend on $x^1$ (they are invariant by 
$\xiv$), rank$(G^0_{ab})=P^a_{\ a}$, rank$(G^1_{ab})=n-P^a_{\ a}$ and $G^{0\ c}_{a}G^1_{cb}=0$.

{\it (ii)} $A$ and $B$ are functions satisfying $\d_{x^1}A=\alpha+\beta$, 
$\d_{x^1}B=\alpha-\beta$. Furthermore the square root $\Su$ is given in these coordinates by 
\be
S_{ab}=e^AG^0_{ab}(x^i)-e^BG^1_{ab}(x^i). 
\label{coordinate-free2}
\ee
\label{adapted}
\end{prop}
\P As is known, local coordinates such that $\xiv=\d/\d x^1$
can always be chosen in a neighbourhood 
of any point in which $\xiv$ does not vanish. In this local coordinate 
system equations 
(\ref{semi-conformal}) become
\be
\fr{\d}{\d x^1}(\G+\Su)=(\alpha+\beta)(\G+\Su),\ \ \ \fr{\d}{\d x^1}(\G-\Su)=(\alpha-\beta)(\G-\Su),
\ee
which can be explicitly integrated giving in components
\bnr
\rmg_{ab}(x^1,x^i)+S_{ab}(x^1,x^i)=2G^0_{ab}(x^i)\exp\left[\int^{x^1}_0ds(\alpha(s,x^i)+\beta(s,x^i))\right],\\
\rmg_{ab}(x^1,x^i)-S_{ab}(x^1,x^i)=2G^1_{ab}(x^i)\exp\left[\int^{x^1}_0ds(\alpha(s,x^i)-\beta(s,x^i))\right].\ \ 
\enr
Addition and subtraction of these equations leads to (\ref{coordinate-free1})-(\ref{coordinate-free2})
with the obvious definitions for $A$ and $B$. 
 $G^0_{ab}$ and $G^1_{ab}$ are proportional to $P_{ab}$ and $\Pi_{ab}$ respectively from what we deduce the stated properties
about their rank and product at once.
\N 

\medskip
\noindent

We finish this section pointing out that it is possible to define {\em conserved quantities and currents} 
for bi-conformal 
vector fields in a similar way as it has been done with other symmetry 
transformations
(see \cite{KERR-SCHILD,PIII} for further details about this).

\section{Lie algebras of bi-conformal vector fields}
\label{LIE}
In this section we will settle under what circumstances bi-conformal vector fields give rise to a subalgebra
 of $\vecX(V)$ and derive some of its basic properties. 
For a fixed square root $\Su$ of the metric $\G$ we denote by $\Gl(\Su)$ the set of bi-conformal vector fields 
whose differential condition (\ref{fundamental}) involves 
$\Su$.  Bi-conformal vector fields belonging 
to $\Gl(\Su_1)$ and $\Gl(\Su_2)$ for two different square roots $\Su_1$ and $\Su_2$ 
are necessarily conformal motions of the metric $\G$ as we are going to prove now. A lemma is needed first.
\begin{lem}
 If $\Su_1$ and $\Su_2$ are square roots of the metric $\G$ such that $\Su_1\times\Su_2=\l\G$
 then $\Su_1=\pm\Su_2$.
\label{S1=S2}
\end{lem} 
\P From the assumptions we easily get 
$$
\Su_1\times\Su_2\times\Su_2=\l\Su_2\Rightarrow \Su_1=\l\Su_2,
$$
and using now $\Su_1\times\Su_1=\Su_2\times\Su_2=\G$ we conclude that $\l^2=1$.\N
\begin{prop}
For two nonproportional square roots $\Su_1$ and $\Su_2$, $\Gl(\Su_1)\cap\Gl(\Su_2)$ is a set of conformal Killing vector
fields.
\label{derivconforme}
\end{prop}
\P The existence of a non vanishing vector field $\xiv$ belonging to 
$\Gl(\Su_1)\cap\Gl(\Su_2)$ entails the relation
\be
\fl\lie\G=\alpha_1\G+\beta_1\Su_1=\alpha_2\G+\beta_2\Su_2\Rightarrow\ 
(\alpha_1-\alpha_2)\G=\beta_2\Su_2-\beta_1\Su_1,
\label{igualdad}
\ee    
Let us assume first that $\beta_1\neq 0$ and $\beta_2\neq 0$. Performing the left inner product and the right inner product
of the last equation with $\Su_1$ we get respectively
$$
\beta_1\G-\beta_2\Su_1\times\Su_2=(\alpha_2-\alpha_1)\Su_1,\ \beta_1\G-\beta_2\Su_2\times\Su_1=(\alpha_2-\alpha_1)\Su_1,
$$
 whose subtraction yields
$$
\beta_2(\Su_1\times\Su_2-\Su_2\times\Su_1)=0\Rightarrow\Su_1\times\Su_2=\Su_2\times\Su_1
$$
On the other hand squaring each member of (\ref{igualdad}) and using $\Su_2\times\Su_2=\G$ we obtain 
$$
-2\beta_1\beta_2\Su_1\times\Su_2=((\alpha_1-\alpha_2)^2-\beta_1^2-\beta_2^2)\G.
$$
Hence, $\Su_1\times\Su_2$ is proportional to the metric tensor which is only 
possible if $\Su_1=\pm\Su_2$ as we proved
in lemma \ref{S1=S2} contradicting the statement of the proposition. 
Thus some of the functions $\beta_1$ or $\beta_2$ must
vanish which implies according to equation (\ref{igualdad}) that either 
$\Su_1=\pm\G$ and $\Su_2=\pm\G$, or $\beta_1=\beta_2=0$,
$\xiv$ being in all of these possibilities a conformal Killing vector.\N

The main conclusion of the previous proposition is the non-existence 
of a nontrivial 
proper bi-conformal vector field constructed from two different square roots $\Su_1$ and $\Su_2$, or in other words 
$\xiv$ cannot leave invariant two different pairs of complementary projectors 
$\{P^a{}_{b},\Pi^a{}_{b}\}$ and $\{P'^a{}_{b},\Pi'^a{}_{b}\}$  (unless it is a conformal motion).

Our next result proves that $\Gl(\Su)$ is a Lie algebra for any square root $\Su$.
\begin{prop}
The set of vector fields in $\Gl(\Su)$ with $\Su$ a square root of the 
metric is a Lie subalgebra of the Lie algebra $\vecX(V)$.
\label{lie-algebra}
\end{prop}
\P We must show that the linear combination and Lie bracket of any pair of vectors $\xiv_1$ and $\xiv_2$ satisfying equations 
(\ref{fundamental}) for a fixed square root $\Su$ is also a solution of this same pair of equations for different gauges.
 Clearly the gauge functions must depend on the chosen 
bi-conformal vector field $\xiv$ in a precise fashion and we will write this dependence as 
$\alpha_{\xiv}$, $\beta_{\xiv}$. After a simple calculation, we get that $\xiv_1+\xiv_2$ and $[\xiv_1,\xiv_2]$ are 
bi-conformal vector fields with gauge functions given by
\bnr
\alpha_{[\xiv_1,\xiv_2]}=\liea\alpha_{\xiv_2}-\lieb\alpha_{\xiv_1},\ \ 
\beta_{[\xiv_1,\xiv_2]}=\liea\beta_{\xiv_2}-\lieb\beta_{\xiv_1},\\ 
\alpha_{\xiv_1+\xiv_2}=\alpha_{\xiv_1}+\alpha_{\xiv_2},\ \ \beta_{\xiv_1+\xiv_2}=\beta_{\xiv_1}+\beta_{\xiv_2}.
\enr
\N
\begin{coro}
The subspace $\Gl(\Su_1)\cap\Gl(\Su_2)$ is a Lie subalgebra of conformal Killing vectors for
every pair of square roots $\Su_1$ and $\Su_2$.
\label{intersection}
\end{coro}
\P This is a consequence of proposition \ref{lie-algebra} and the fact that the intersection of two Lie algebras is
a Lie subalgebra.\N

We analyze next some properties of the Lie algebra $\Gl(\Su)$ which might shed 
some light on the nature of bi-conformal vector fields. First of all it is 
clear that $\Gl(\Su)$ may 
sometimes consist only on the trivial solution and thus $\Gl(\Su)=\{\vec{\bf 0}\}$ which 
is a case of no interest. When $\Gl(\Su)\neq\{\vec{\bf 0}\}$ 
an important question concerns their 
dimensionality as vector spaces. As happens with other known transformations in 
Differential Geometry, this dimension can be either finite or infinite depending on
whether one is able to get a normal system of partial differential equations out of 
(\ref{fundamental}). Here normal means that the first derivatives of a well-defined set
of unknowns can be isolated in terms of themselves, see \cite{EISENHARTII} for a discussion and proof of this result
and section \ref{highest}.  
Nonetheless the search of a normal system 
to discard the existence of infinite dimensional Lie algebras can
sometimes be bypassed if we state under what circumstances $\xiv$ and $\rho\xiv$ 
are bi-conformal vector fields for a function $\rho$. We settle
 this issue in the following proposition.
\begin{prop}
If both vector fields $\xiv$ and $\rho\xiv$ are in $\Gl(\Su)$ for some non-constant 
function $\rho$ then $\O$ is either a 1-form with 
$\O\propto\bxi$ or a 2-form $\O\propto\bxi\wedge d\rho$ where $\O$ is the simple form generating the 
square root $\Su$.
\label{PROPORTIONALITY} 
\end{prop}
\P To prove this we will study when the vector field $\rho\xiv$ solves 
(\ref{fundamental}) given that $\xiv$ does, being $\rho$ a smooth function. 
Substitution of this in (\ref{fundamental}) yields ($\rho_a=\d_a\rho$)
\bea
\rho_a\xi_b+\rho_b\xi_a=(\bar{\alpha}-\rho\alpha)\rmg_{ab}+
(\bar{\beta}-\rho\beta)S_{ab}\label{1tandaa}\\
\rho_a\xi_cS^{c}_{\ b}+\rho_b\xi_cS^{c}_{\ a}=(\bar{\beta}-\rho\beta)\rmg_{ab}
+(\bar{\alpha}-\rho\alpha)S_{ab},\ \
\label{1tanda}
\eea
where $\bar{\alpha}$, $\bar{\beta}$ are the gauges of $\rho\xiv$.
Multiplying by $S^{b}_{\ c}$ we get after a suitable relabelling of indexes
\bea 
\rho_a\xi_cS^{c}_{\ b}+\xi_a\rho_cS^c_{\ b}=(\bar{\alpha}-\rho\alpha)S_{ab}+
(\bar{\beta}-\rho\beta)\rmg_{ab},\label{2tandaa}\\
\rho_a\xi_b+\xi_eS^e_{\ a}\rho_cS^c_{\ b}=(\bar{\beta}-\rho\beta)S_{ab}+
(\bar{\alpha}-\rho\alpha)\rmg_{ab}.
\label{2tanda}
\eea 
Subtracting  (\ref{2tandaa}) from (\ref{1tanda}), and (\ref{2tanda}) from (\ref{1tandaa}) we get
\be
\xi_a\rho_cS^c_{\ b}-\rho_b\xi_cS^c_{\ a}=0,\ \ \ \ \rho_cS^c_{\ b}\xi_eS^e_{\ a}-\rho_b\xi_a=0,
\ee
which can be reduced to a couple of linear homogeneous systems 
 by contracting with $\rho_cS^{cb}$ and $\xi^a$ both relations
\be
\fl\left.\begin{array}{c}
      \xi_a(\rho_c\rho^c)-(\rho_cS^{cb}\rho_b)S^e_{\ a}\xi_e=0\\
      \rho_c\rho^c\xi_eS^e_{\ a}-\rho^cS_{cb}\rho^b\xi_a=0\end{array}\right\},\ \ 
\left.\begin{array}{c}
          (\xi^a\xi_a)\rho_cS^c_{\ b}-\rho_b(\xi_cS^{c}_{\ a}\xi^a)=0\\
         (\xi^eS_{ec}\xi^c)\rho_qS^q_{\ b}-(\xi_c\xi^c)\rho_b=0\end{array}\right\}.\ \ 
\ee     
The fulfillment of both systems implies that $\xi_bS^b_{\ c}=\epsilon\xi_c$ and 
$\rho_bS^b_{\ c}=\epsilon\rho_c$ with $\epsilon^2=1$. 
These  conditions entail a further relation arising from 
(\ref{1tanda}) and (\ref{2tanda}) between the barred and unbarred gauges given by 
$\bar{\alpha}-\rho\alpha=\epsilon(\bar{\beta}-\rho\beta)$. From this we deduce that 
(\ref{1tandaa}) takes the form 
$$
\rho_a\xi_b+\rho_b\xi_a=(\bar{\alpha}-\rho\alpha)(\rmg_{ab}+\epsilon S_{ab}).
$$ 
Since the tensors $(\rmg_{ab}+\epsilon S_{ab})/2$ are projectors 
(either $P_{ab}$ or $\Pi_{ab}$) they can be 
thought of as the metric tensor of a certain subspace (the subspace generated by the vectors forming the simple form 
generating $\Su$) meaning that $d\rho\otimes\bxi+\bxi\otimes d\rho$ can only 
be algebraically such a projector if either the subspace is one-dimensional generated by $\bxi$ with 
$\bxi\propto d\rho$ or two dimensional with $d\rho$ and $\bxi$ as generators. If $\epsilon=1$ this 
projector is $P_{ab}$ so $S_{ab}$ is the superenergy tensor
of a simple form proportional to $\bxi$ in the 
one-dimensional case or to $\bxi\wedge d\rho$ in the two dimensional case
 (see considerations coming after equation (\ref{losproyectores})).
The discussion for the case $\epsilon=-1$ is similar replacing $P_{ab}$ by $\Pi_{ab}$.\N

\noindent
{\bf Remark.} This result can be extended to include the dual of $\O$,
which is either a $(n-1)$-form or a $(n-2)$-form, by means of the property
$$
\T\{*\O\}=(-1)^{n-1}\fr{|\det(\G)|}{\det(\G)}\T\{\O\}
$$
holding for any superenergy tensor defined by equation (\ref{superenergy}).

 We see as an outcome of this proposition that the cases with  
$P^a_{\ a}=1,2,n-1,n-2$ may contain infinite-dimensional Lie algebras $\Gl(\Su)$.
This is what one should expect given the conformal character of these symmetries in the subspaces
on which $P_{ab}$ and $\Pi_{ab}$ project because either of these subspaces
will be of dimensions less or equal than two whenever $\Su$ is built up from a 
1-form or a 2-form (or their duals) and the conformal group in these dimensions can be 
of infinite dimension as is well known \cite{YANO}.   

Another interesting result coming up 
from the proof of this proposition is that, in the case of $\Su=T\{\O\}$ 
for a 1-form $\O$, this 1-form must be proportional to ${\bxi}\propto d\rho$
from what we conclude that $\xiv$ is irrotational. Indeed the converse of this
statement is also true.      
\begin{prop}
Let $\xiv$ be a  bi-conformal vector field such that $\Su\propto\T\{\bxi\}$. Then 
$\rho\xiv$ is also a bi-conformal vector field iff $d\rho\wedge{\bxi}=0$.
\label{PARTIAL}
\end{prop} 
\P The ``if'' is a particular case of proposition \ref{PROPORTIONALITY} 
so we are only left with the ``only if'' implication. Under the hypotheses of 
this proposition the tensor $S_{ab}$ takes the form (equation (\ref{superenergy}) with
$\O=\bxi$)
$$
S_{ab}=-\rmg_{ab}+2\xi^{-2}\xi_a\xi_b,\ \ \xi^2\equiv\xi_a\xi^a. 
$$
If (\ref{fundamental}) is assumed, the first equation of the couple becomes
\be
\nb_a\xi_b+\nb_b\xi_a=(\alpha-\beta)\rmg_{ab}+2\xi^{-2}\beta\xi_a\xi_b .
\label{nono}
\ee
 Furthermore the second equation of (\ref{fundamental}) is a consequence of this because
\bnr
\lie\xi_a=\lie(\rmg_{ap}\xi^p)=(\alpha\rmg_{ab}+\beta S_{ab})\xi^b=
(\alpha+\beta)\xi_a \,\, \Longrightarrow \\
\Longrightarrow \,\, \lie S_{ab}=-\lie\rmg_{ab}+2
\lie\left(\xi^{-2}\xi_a\xi_b\right)=\\
\fl=-\alpha\rmg_{ab}-\beta S_{ab}-2\xi^{-2}(\alpha+\beta)\xi_a\xi_b+
4\xi^{-2}(\alpha+\beta)\xi_a\xi_b=
\alpha\left(-\rmg_{ab}+2\xi^{-2}\xi_a\xi_b\right)+\beta\rmg_{ab},
\enr 
so we only need to care about (\ref{nono}). The 1-form $\rho\xi_a$ 
fulfills the equation
$$
\nb_a(\rho\xi_b)+\nb_b(\rho\xi_a)=
(\bar{\alpha}-\bar{\beta})\rmg_{ab}+2\bar{\beta}
(\rho\xi)^{-2}\rho\xi_a\rho\xi_b =
(\bar{\alpha}-\bar{\beta})\rmg_{ab}+2\xi^{-2}\bar{\beta}\xi_a\xi_b
$$  
as long as $\gamma\xi_b=\d_b\rho$ for some smooth function $\g$. 
The gauges $\bar{\alpha}$ and $\bar{\beta}$ are given then by 
 $\bar{\beta}=\xi^2\g+\beta\rho$ and $\bar{\alpha}=\xi^2\g+\alpha\rho$.
\N

\section{Normal system and maximal Lie algebras of bi-conformal vector fields}
\label{highest}
We start now to tackle the integrability conditions of equations (\ref{fundamental}) for a fixed Lorentz tensor as 
well as the greatest dimension of the vector space $\Gl(\Su)$ when it happens to be finite dimensional.
First of all, we give an overview of the procedure to be followed which is quite similar
to the one used to find out the first integrability conditions and the largest dimension 
of Lie algebras for isometries, conformal motions or collineations 
(see e.g. \cite{YANO} for an account of this). The aim is to rewrite the equations satisfied by $\xiv$, by using 
successive derivatives and identities, in {\em normal} form, that is to say, 
such that one can identify a definite set of unknowns,
say $Z_A$, for which the equations become a set of first order PDEs 
like (\ref{general-normal-form}), see below, with all 
the derivatives isolated. In our case, this normal system, will actually be 
linear and homogeneous, in case it exists, as we are going to prove.
    
 One starts with the differential 
conditions fulfilled by each generator of the symmetry under study written generically
in the form, 
\be
\Phi_I(\xiv,\nb\xiv,\dots,\phi_1,\nb\phi_1,\dots)=0,     
\label{condition}
\ee
where $\phi_1,\dots$ are some scalar functions accounting for the gauges of the symmetry and the 
index $I$ denotes the whole set of tensor indexes which appear in the differential conditions. 
These equations must be differentiated a number of times resulting in new algebraic
equations involving the variables appearing in (\ref{condition}) plus higher
derivatives of them (we now use the subindex $I_{k}$ to gather  
the resulting new indexes) 
\be
\Phi^{(k)}_{I_{k}}(\xiv,\nb\xiv,\nb^2\xiv,\dots,\phi_1,\nb\phi_1,\nb^2\phi_1,\dots)=0,\ k=1,2,\dots.
\label{derivatives}
\ee 
 From these equations one wants to get another set which contains the first derivatives
of the {\it system variables} isolated in terms of themselves and the manifold data. 
To do this one may need to include higher derivatives of the initial
 variables $\xiv$, $\phi_1$, $\dots$ as new independent variables so these definitions
will be part of the normal system. Hence the normal set of equations will in general look like
\be
D_aZ_A=f_{aA}(x,Z),
\label{general-normal-form}
\ee
where $x\!\!=(x^1,\dots,x^{n})$ is the chosen local coordinate system, 
$Z=\{Z_A\}=(Z_1(x),\dots,Z_m(x))$ denotes the complete set of system 
variables, $D_a$ is a differential operator and $f_{aA}$ are functions 
depending on the coordinates and the system variables. 
A normal system of equations can only be achieved for finite-dimensional 
Lie groups. An important point regarding this calculation is that
the chain of equations (\ref{derivatives}) used to get the normal 
system may give rise 
to constraints between the system variables $Z_A$ in the form
\be
L_C(x,Z)=0,\ \ \ C=1,\dots,q.
\label{general-constraint}
\ee
These constraints arise when some of the equations of the above chain (or some linear combination of them) do not 
contain derivatives of the system variables. 
 
The first integrability conditions are new relations between the 
system variables coming from the compatibility of the anti-symmetrized second derivatives 
$D_{[a}D_{b]}$ and the normal system (\ref{general-normal-form}) 
\bea
 2D_{[b}D_{a]}Z_A=D_bf_{aA}(x,Z)+\sum_{C}\fr{\d f_{aA}(x,Z)}{\d Z_C}f_{bC}(x,Z)-\nonumber\\
-D_af_{bA}(x,Z)-\sum_{C}\fr{\d f_{bA}(x,Z)}{\d Z_C}f_{aC}(x,Z),
\label{cross-derivative}
\eea
which can be arranged in the same form as (\ref{general-constraint}) 
where some identity for $D_{[a}D_{b]}$ must be used in this last step (for instance if $D_a$ is 
the covariant derivative, the Ricci identity (\ref{ricci})). 
The constraints (\ref{general-constraint}) themselves must be differentiated 
(propagated)
\be
D_aL_C(x,Z)+\sum_{B}\fr{\d L_C(x,Z)}{\d Z_B}D_bZ_B=0,
\label{constraint-derivative}
\ee
getting new relations involving the system variables and the data which 
are the first integrability conditions coming from
the constraints. They must be added to the set arising from the commutation of the derivatives. The 
whole set of first integrability conditions will be identically satisfied if the $(V,\G)$ we deal with is 
maximal with respect to the symmetry under study. The fulfillment of the 
maximal integrability will pose certain geometric conditions which characterize 
these maximal spaces. In maximal 
spaces arising from a normal form, the largest dimension of the Lie algebra of vector fields satisfying 
(\ref{condition}) is achieved, and this largest dimension is given by 
the total number of system variables $m$ minus the number of independent constraints $q$ found in 
(\ref{general-constraint}). The set of solutions 
of (\ref{condition}) is a subspace of $\vecX(V)$ which thus depends linearly on $m-q$ arbitrary constants.  
If the first integrability conditions are not identically satisfied, we must carry on the above described procedure 
but now applied to the first integrability conditions obtaining a chain 
of equations of the same type
\be
\Xi^j(x,Z)=0,\ \ j=q+1,\dots 
\label{further-derivatives}
\ee 
This new set of integrability conditions impose further geometric constraints for each $j$ 
(we are assuming that all equations in (\ref{further-derivatives}) are algebraically independent). 
If there exists a value of $j$ such that the corresponding condition is identically 
satisfied (or we are able to settle the geometric conditions for this to happen) then the solution of the differential 
conditions is a Lie subalgebra of vector fields of dimension $m-j$. On the other hand, if the number of linearly independent 
equations in (\ref{further-derivatives}) and (\ref{general-constraint}) 
is greater or equal than $m$ then we get a homogeneous linear system
for the system variables with no solution other than the trivial one.

\subsection{Normal system}
In what follows, we are going to construct the normal system form for bi-conformal vector fields ---in the cases this is
 possible--- thereby obtaining also the largest dimension of some finite dimensional $\Gl(\Su)$ by following 
the above outlined procedure. This will in particular allow us to prove rigorously that the 
cases identified in the previous section are 
the only ones with a feasible infinite dimension. This is a rather long
calculation and only its main excerpts are shown. 
To start with, we rewrite (\ref{fundamental}) (or 
(\ref{semi-conformal})) in terms of the projectors 
$P_{ab}$ and $\Pi_{ab}$ as 
\bea
\lie P_{ab}=\phi P_{ab}, \ \ 
\lie \Pi_{ab}=\chi \Pi_{ab}, \label{1y2}\\
\Rightarrow \ \ \lie P^{a}_{\ b}=0, \ \ \lie \Pi^{a}_{\ b}=0, \nonumber\\ 
\Rightarrow \ \ \lie P^{ab}=-\phi P^{ab}, \lie \Pi^{ab}=-\chi \Pi^{ab} \nonumber
\eea
where $\phi=(\alpha+\beta)$ and $\chi=(\alpha-\beta)$ and the second 
pair is (\ref{mixed-inv}). The following well known formulae in differential geometry are needed
(see \cite{YANO} for an account of them)
\bea
\lie\g^a_{bc}=\fr{1}{2}\rmg^{ae}\left[\nabla_b(\lie\rmg_{ce})+
\nabla_c(\lie\rmg_{be})-\nabla_e(\lie\rmg_{bc})\right],\label{lie-connection}\\
\lie\g^a_{bc}=\nabla_b\nabla_c\xi^a+\xi^dR^a_{\ cdb},\label{lie-xi}\\
\hspace{-1cm}\nabla_c\lie T^{a_1\dots a_p}_{\ b_1\dots b_q}-
\lie\nabla_cT^{a_1\dots a_p}_{\ b_1\dots b_q}=\nonumber\\
=-\sum_{j=1}^p(\lie\g^{a_j}_{cr})T^{\dots a_{j-1}ra_{j+1}\dots}_{\ b_1\dots b_q}+
\sum_{j=1}^q(\lie\g^r_{cb_j})T^{a_1\dots a_p}_{\dots b_{j-1}rb_{j+1}\dots},
\label{lie-conmmutation}\\
\lie R^d_{\ cab}=\nabla_a(\lie\g^d_{\ bc})-\nabla_b(\lie\g^d_{ac})
\label{lie-curvature},
\eea
where $\g^a_{bc}$ is the Levi-Civita connection of the metric
$\rmg_{ab}$ and $R^d_{\ cab}$  its curvature tensor.
Our convention for the Riemann tensor is such that the Ricci identity is
\be
\nb_a\nb_bu_c-\nb_b\nb_au_c=u_dR^d_{\ cba}.
\label{ricci}
\ee
 The Lie derivative of the connection can be worked out at once since
we know $\lie\G$. In terms of the quantities defined in (\ref{1y2}) it becomes
($\phi_b\equiv\d_b\phi,\ \chi_b\equiv\d_b\chi$)
\be
\fl\lie\g^a_{bc}=\fr{1}{2}(\phi_bP^a_{\ c}+\phi_cP^a_{\ b}-\phi^aP_{bc}+
\chi_b\Pi^a_{\ c}+
\chi_c\Pi^a_{\ b}-\chi^a\Pi_{cb}+(\phi-\chi)M^a_{\ bc}),
\label{3}
\ee
where we have defined the tensor
\bea
\hspace{-1.4cm} M^a_{\ bc}=\nabla_bP^a_{\ c}+\nabla_cP^a_{\ b}-\nabla^aP_{cb} 
=-(\nabla_b\Pi^a_{\ c}+\nabla_c\Pi^a_{\ b}-\nabla^a\Pi_{bc}),\label{4y5}\\
 M^a_{\ bc}=M^a_{(bc)},\ M^a_{\ ac}=0,\nonumber
\eea
coming the second equality in (\ref{4y5}) from 
$\nabla_cP_{ab}=-\nabla_c\Pi_{ab}$. $M^a_{\ bc}$ can be equally 
defined from the square root $\Su$ as
$$
M^a_{\ bc}=\fr{1}{2}\left(\nabla_bS^a_{\ c}+\nabla_cS^a_{\ b}-\nabla^aS_{cb}\right).
$$
Observe also that
$$
P^{ca}M_{acb}=P^{ca}(\nb_cP_{ab}+\nb_bP_{ac}-\nb_aP_{bc})=
P^{ca}\nb_bP_{ac}=0
$$
as follows from $0=\nb_b(P^{ca}P_{ac})$, and analogously 
$$
\Pi^{ab}M_{abc}=0 .
$$
Using (\ref{lie-conmmutation}) and (\ref{3}) we can calculate the 
Lie derivative of $M_{abc}$
\bea
\lie M_{abc}=\phi M_{abc}+(\chi-\phi)P_{ap}M^{p}_{\ bc}-       
P_{bc}\Pi_{ap}\phi^p+\Pi_{cb}P_{ap}\chi^p\nonumber\\
=\chi M_{abc}+(\phi-\chi)\Pi_{ap}M^p_{\ bc}-       
P_{bc}\Pi_{ap}\phi^p+\Pi_{cb}P_{ap}\chi^p\label{9},
\eea
where the identity 
$\phi M_{abc}+(\chi-\phi)P_{ap}M^p_{\ bc}=\chi M_{abc}+(\phi-\chi)\Pi_{ap}M^p_{\ bc}$
must be used to get the last equality. 
The trace of (\ref{9}) can be split in two if we contract with $P^{cb}$ and $\Pi^{cb}$ 
respectively getting
\bea
\lie(M_{abc}P^{bc})+(\phi-\chi)P_{ap}M^{p}_{\ bc}P^{bc}=-p\Pi_{ap}\phi^p\nonumber\\
\lie(M_{abc}\Pi^{bc})+(\chi-\phi)\Pi_{ap}M^p_{\ bc}\Pi^{bc}=-(n-p)P_{ap}\chi^p\nonumber,
\eea
(recall that $p=P^{a}_{\ a})$. 
A further simplification of this arises if we realize the property 
$0=P_{ap}M^{p}_{\ cb}P^{cb}=\Pi_{ap}M^p_{\ cb}\Pi^{cb}$ so the last couple of 
equations yields
\be
\lie E_a=-p\Pi_{ap}\phi^p,\ \ \ \lie W_a=(p-n)P_{ap}\chi^p,
\label{18}
\ee
where 
$$
E_a\equiv M_{acb}P^{cb},\ \ W_a\equiv-M_{acb}\Pi^{cb},\ \Pi_{ac}E^c=E_a,\  P_{ac}W^c=W_a,\ 
0=P^{ab}E_b=\Pi^{ab}W_b.
$$
 Now we can use equation (\ref{lie-curvature}) to work out $\lie R^d_{\ cab}$ obtaining
\bea
\fl\lie R^d_{\ cab}=P^d_{\ [b}\nb_{a]}\phi_c+P_{c[a}\nb_{b]}\phi^d+\Pi^d_{\ [b}\nb_{a]}\chi_c+\Pi_{c[a}\nb_{b]}\chi^d+
\phi_{[b}\nb_{a]}P^d_{\ c}+\phi_c\nb_{[a}P^d_{\ b]}\nonumber\\
\fl+\phi^d\nb_{[b}P_{a]c}+\chi_{[b}\nb_{a]}\Pi^d_{\ c}+\chi_c\nb_{[a}\Pi^d_{\ b]}+\chi^d\nb_{[b}\Pi_{a]c}+
\nb_{[a}[M^d_{\ b]c}(\phi-\chi)].
\label{lie-riemann}
\eea
Multiplying here by $P^a_{\ d}$, using algebraic properties and $\nb_b\phi_c=\nb_c\phi_b$
we get
\bea
\fl2\lie(P^a_{\ d}R^d_{\ cab})=-\nb_c(\Pi^a_{\ b}\phi_a)-\nb_b(\Pi^a_{\ c}\phi_a)-
(\nb_cP^a_{\ b}+\nb_bP^a_{\ c}+P^{pa}\nb_pP_{bc})\phi_a+\nonumber\\
\fl+\left(2-p\right)\nb_b\phi_c-P^{ad}\nb_a\phi_dP_{bc}-P^{ad}\nb_a\chi_d\Pi_{cb}
+\fr{1}{2}(\phi_bE_c+\phi_cE_b-\chi_bE_c-\chi_cE_b)-\nonumber\\
\fl-P^{ad}\chi_d\nb_a\Pi_{cb}+P^a_{\ d}\{\nb_a[(\phi-\chi)M^d_{\ bc}]-\nb_b[(\phi-\chi)M^d_{\ ac}]\}.
\label{45}
\eea
A further contraction of this equation with $P^{cb}$ yields
\be
2(1-p)P^{cb}\nb_b\phi_c=2\lie R^0+2\phi R^0+(\chi-\phi)W^0,
\label{47}
\ee
where $R^0=P^{cb}P^{ar}R_{rcab}$ and $W^0=P^{ar}(\nb_aM_{rbc}-\nb_bM_{rac})P^{cb}$. After substitution of 
(\ref{47}) into (\ref{45}) we realize that,
in order to get an expression with the derivatives of $\phi_b$ and $\chi_b$ isolated, 
the only  terms which require of a further treatment are
$$
-\nb_c(\phi_r\Pi^r_{\ b})+\nb_b(\phi_r\Pi^r_{\ c})-P^{ad}\nb_a\chi_d\Pi_{cb},
$$
which can be worked out by derivating the second equation of (\ref{18}) and 
using (\ref{lie-conmmutation}). The result of such calculation is
\bea
\fl\nb_c(\Pi_{ar}\phi^r)=-\fr{1}{p}[\lie(\nb_cE_a)-\fr{1}{2}(E_r\phi^r)P_{ac}-
\fr{1}{2}(E_r\chi^r)\Pi_{ac}+\chi_{(a}E_{c)}-\fr{1}{2}(\chi-\phi)E_rM^r_{\ ca}],
\nonumber\\ \label{derivada1}\\
\fl\nb_c(P_{ar}\chi^r)=\fr{-1}{n-p}[\lie(\nb_cW_a)-\fr{1}{2}(W_r\chi^r)\Pi_{ac}
-\fr{1}{2}(W_r\phi^r)P_{ac}+\phi_{(c}W_{a)}-\fr{1}{2}(\chi-\phi)W_rM^r_{\ ca}].
\nonumber\\ \label{derivada2} 
\eea
Equation (\ref{derivada1}) provides us the answer for the terms 
$-\nb_c(\f_r\Pi^r_{\ b})+\nb_b(\f_r\Pi^r_{\ c})$ whereas a further manipulation of 
(\ref{derivada2}) yields
\bea
P^{ad}\nb_a\chi_d=\nb_a(P^{ad}\chi_d)-\fr{1}{2}(E^d-W^d)\chi_d\ \ \Longrightarrow\nonumber\hspace{3.7cm}\\
\fl P^{ad}\nb_a\chi_d=\fr{-1}{n-p}\lie(\nb_aW^a)+W_r\chi^r+\fr{p-2}{2(n-p)}W_r\phi^r
-\fr{1}{2}E^r\chi_r-\fr{1}{n-p}\phi\nb_rW^r.\ \ \ \ \label{ide2}
\eea 
Therefore substitution of (\ref{ide2}), (\ref{derivada1}) and (\ref{47}) into 
(\ref{45}) results, after some tedious algebra, in
\bea
\fl\nb_c\phi_b=\fr{1}{2-p}
\lie\left[2P^{ad}R_{dcab}-\fr{2}{p}\nb_{(c}E_{b)}+\fr{R^0}{1-p}P_{bc}-\fr{1}{n-p}\nb_rW^r\Pi_{cb}\right]
+\nonumber\\
\fl+\left(\fr{1}{2p}E_r\chi^r-\fr{1}{2(n-p)}W_r\phi^r+\fr{1}{2-p}W_r\chi^r\right)\Pi_{cb}-
\fr{1}{p}\chi_{(c}E_{b)}-\fr{1}{2-p}\phi_{(b}E_{c)}+\nonumber\\
\fl+\left(\fr{2}{2-p}\nb_{(c}P^r_{\ b)}+\fr{1}{p(2-p)}E^rP_{cb}\right)\phi_r+\fr{(\phi_r-\chi_r)}{2-p}
(-2P^{dr}\nb_{(b}P_{c)d}+2P^{pr}\nb_pP_{bc})+\hspace{1.5cm}\nonumber\\
\fl+\fr{(\chi-\phi)}{2-p}(P^{ad}\nb_aM_{dbc}+\nb_bP^{ad}\nb_cP_{ad}+
\fr{1}{p}E_rM^r_{\ cb}+\fr{W^0}{2(1-p)}P_{bc}+\fr{1}{n-p}\nb_rW^r\Pi_{cb}).
\label{norm}
\eea
This equation has a counterpart obtained by means of the replacements 
$\phi\leftrightarrow\chi$, $P_{ab}\leftrightarrow \Pi_{ab}$ and $p\leftrightarrow n-p$ 
which we shall omit for the sake of brevity.
Equation (\ref{norm}) and its counterpart together with the following ones 
\be
\eqalign{
\d_a\phi=\phi_a\nonumber\\ 
\d_a\chi=\chi_a\nonumber\\ 
\nb_a\xi_b=\Psi_{ab}+\fr{1}{2}(\phi P_{ab}+\chi\Pi_{ab}),\ \ \Psi_{[ab]}=\Psi_{ab}\nonumber\\
\nabla_b\Psi_{ca}=\xi_dR^d_{\ bca}+\phi_{[c}P_{a]b}+\chi_{[c}\Pi_{a]b}+(\phi-\chi)\nb_{[c}P_{a]b},}
\label{normal-set}
\ee
comprise the normal system,
where the third equation is the first of (\ref{fundamental}) written in terms of the system variables and
the formula for $\nabla_b\Psi_{ca}$ is (\ref{lie-xi}) with the Lie derivatives of
the connection replaced by the values given in (\ref{3}).
\subsection{Constraint equations}
The variables appearing in equations (\ref{norm}), its counterpart, and (\ref{normal-set})
 are not independent because, as we are going to show next, they must fulfill
a certain set of constraints reducing the effective number of them. To realize the existence of such constraints, 
let us review the procedure we have followed to derive the normal system out of the original differential conditions.
 We started with the differential conditions (\ref{1y2}), differentiated them obtaining (\ref{9}) and the fourth equation of 
(\ref{normal-set}),
 and then we calculated the second covariant derivative of the differential conditions 
(which essentially is (\ref{lie-riemann}))
yielding (\ref{norm}) and its partner after some algebraic manipulations involving the differentiation of 
both equations in (\ref{18}). 
Therefore the equations which play the role of the chain written in (\ref{derivatives}) are 
\begin{center}
\begin{itemize}
\item differential conditions . . . . . . . . . . . . . . . . . . (\ref{1y2}).
\item linear combination of first derivatives . . .(\ref{lie-xi}), (\ref{18}), $\phi_a=\d_a\phi$, $\chi_a=\d_a\chi$. 
\item linear combination of second derivatives . . . . . (\ref{45}), (\ref{derivada1}), (\ref{derivada2}), (\ref{ide2}),
\end{itemize} 
\end{center}
 because these and only these equations are used to get the normal system (\ref{norm}) and (\ref{normal-set}). 
Nonetheless, some
of the previous equations (or suitable linear combinations) do not contain derivatives of the system variables when 
expanded in terms of them which means that they are constraints of type (\ref{general-constraint}). These are
\bea
\xi^c\nabla_cS_{ab}+\Psi_{ac}S^c_{\ b}+\Psi_{bc}S^c_{\ a}=0,\ \ \ S_{ab}=P_{ab}-\Pi_{ab}
\label{CONSTRAINT1}\\
\xi^c\nb_cE_a+\Psi_{ac}E^c=-\fr{1}{2}\chi E_a-\fr{1}{2}(n+p)\Pi_{ap}\phi^p,
\label{CONSTRAINT2}\\
\xi^c\nb_cW_a+\Psi_{ac}W^c=-\fr{1}{2}\phi W_a-\fr{1}{2}(n-p)P_{ap}\chi^p,
\label{CONSTRAINT3}
\eea
where the first constraint is the second of (\ref{fundamental})
(equivalently a linear combination of (\ref{1y2})) and (\ref{CONSTRAINT2}), (\ref{CONSTRAINT3}) are the first and second of 
(\ref{18}) respectively. These constraints must be appended to (\ref{norm}), its partner, and (\ref{normal-set}) and they are 
needed to settle the true number of independent variables.
 In appendix A we will prove that the first expression only entails 
$p(n-p)$ independent equations whereas the next couple contain $n-p$ and $p$ independent equations
respectively. This last statement can be seen for (\ref{CONSTRAINT2}) if we note that $P_a^{\ b}E_b=0$
which means that we have at most as many equations as the dimension of the subspace orthogonal to $P_a^{\ b}$, 
that is, $n-p$. The same reasoning gives $p$ independent 
equations for (\ref{CONSTRAINT3}), whence both equations 
amount for $n-p+p=n$ independent equations as claimed.
\subsection{Maximal spaces} 
The full normal system will always make sense unless either of $2-p$, $1-p$,
$2-n+p$, $1-n+p$ vanishes ($n=\pm p$ must be discarded here since it is only 
possible if either $P_{ab}$ or $\Pi_{ab}$ vanishes). 
We will see in the next subsection that these cases correspond with the ones found in proposition 
\ref{PROPORTIONALITY} but before doing that let us give the sought upper bound for the dimension of $\Gl(\Su)$ 
when there is a normal system. This number is the total number of 
variables appearing in the system (\ref{norm}), its counterpart and (\ref{normal-set}) minus 
the constraints (\ref{CONSTRAINT1}), (\ref{CONSTRAINT2}), (\ref{CONSTRAINT3})\small 
\bnr
\fl\begin{array}{c|c|c|c|c}
  counting   & n  & C_{n,2}  & 2 & 2n \\ \hline
variables & \xi_a & \Psi_{ab} & \phi,\chi & \phi_r,\chi_r
\end{array}\ \ \ \ \ \ \ \ \  
\begin{array}{c|c|c|c}
counting & p(n-p) & n-p & p \\ \hline
constraints & (\ref{CONSTRAINT1})&
(\ref{CONSTRAINT2}) & (\ref{CONSTRAINT3}) 
\end{array}
\enr
\normalsize
Thus an upper bound $N$ for the dimension of $\Gl(\Su)$ is 
$N=n+C_{n,2}+2+2n-p(n-p)-n=\fr{1}{2}[n^2+n(3-2p)+4+2p^2]$. The natural number $N$  
can be rewritten as $(p+1)(p+2)/2+(n-p+1)(n-p+2)/2$. We have thus proven
\begin{theo}
If $p,n-p\notin\{1,2\}$, every Lie algebra of bi-conformal vector fields $\Gl(\Su)$ 
has finite dimension with
$$
dim(\Gl(\Su))\leq\fr{1}{2}(p+1)(p+2)+\fr{1}{2}(n-p+1)(n-p+2)\equiv N.
$$
\label{numero-maximo}
\end{theo}
The right hand side of this inequality is the sum 
of the maximum number of conformal motions in $p$ dimensions plus the same number 
for $n-p$ dimensions suggesting that the maximum dimension of $\Gl(\Su)$ is achieved
when our space can be split in two conformally flat dimensional pieces. 
We can show that this is true and that the maximum dimension can be actually realized.
\begin{prop}
The previously defined number $N$ is the maximum dimension of $\Gl(\Su)$ if $p,\ n-p\notin\{1,2\}$ being
this dimension attained for any $(V,\G)$ whose line element is in local coordinates $\{x^a\}$, $a=1,\dots, n$
\be
ds^2=\phi_1^2(x^a)\eta^0_{\alpha\beta}dx^{\alpha}dx^{\beta}+\phi_2^2(x^a)\eta^1_{AB}dx^Adx^B,
\label{twopieces}
\ee
where $x^{\alpha}=\{x^1,\dots,x^p\}$, $x^A=\{x^{p+1},\dots,x^{n}\}$ are sets of coordinates and
$\bfeta^0$, $\bfeta^1$ flat metrics of the appropriate signatures
depending only on the coordinates $\{x^{\alpha}\}$ and $\{x^{A}\}$ respectively.
\label{canonical-form}
\end{prop}

\noindent
{\bf Remark}. This result is local, valid in any neighbourhood in which
the local coordinates are defined. The maximum number of {\em global} 
bi-conformal vector fields can be obviously less than $N$.

\noindent
\P From equation (\ref{1y2}) is clear that every conformal vector field of $\bfeta^0$
or $\bfeta^1$ is a bi-conformal vector field of $(V,\G)$ because the projectors $\bm{P}$ and 
$\bm{\Pi}$ are given in terms of the flat metrics $\bfeta^0$ and $\bfeta^1$ by 
$$   
\bm{P}=\phi_1^2\bfeta^0,\ \ {\mathbf \Pi}=\phi_2^2\bfeta^1,
$$
whence
$$
\liea\bm{P}=\left(\phi_1^{-2}\liea(\phi_1^2)+\s_1\right){\bm{P}},\ \ \ 
\liea{\mathbf\Pi}=\phi_2^{-2}\liea(\phi_2^2)\, {\mathbf\Pi},
$$
where $\xiv_1$ is any conformal Killing vector of the metric $\bfeta^0$, that is, $\liea\bfeta^0=\s_1\bfeta^0$, 
$\liea\bfeta^1=0$. Similarly we can prove the result for conformal Killing vectors 
of the metric $\bfeta^1$. Thus the line element given by equation (\ref{twopieces}) is maximally 
symmetric for bi-conformal vector fields since there are 
$\fr{1}{2}(p+1)(p+2)$ linearly independent conformal Killing vectors for the metric $\bfeta^0$ and 
$\fr{1}{2}(n-p+1)(n-p+2)$ for the metric $\bfeta^1$ being mutually linearly independent. \N

\noindent
{\bf Remark}. Proposition \ref{canonical-form} has an obvious generalized validity
for the cases where $p$ and/or $n-p$ are 1 or 2. The statement then 
is that {\em every conformal Killing vector of either of the two pieces $\bfeta^0$
or $\bfeta^1$ is a bi-conformal vector field of the space $(V,\G)$}. 

\subsection{Infinite dimensional Lie algebras of bi-conformal vector fields}
As we have already mentioned, the normal system given by the set (\ref{norm}), its partner, and 
(\ref{normal-set}) cannot be defined for some values of the trace $p=P^a_{\ a}$ which is the dimension
 of the subspace onto which $P^a_{\ b}$ projects. Therefore   
 these values of $p$ are linked to the possibility of infinite-dimensional Lie algebras 
of bi-conformal vector fields. Those values were
$$
p=2,\ \ p=1,\ \ p=n-2,\ \ p=n-1,
$$  
in which case (\ref{norm}) and its partner are not defined and there is no
 normal system for the variables $\xi_a$, $\Psi_{ab}$, $\phi$, $\chi$, $\phi_r$ and $\chi_r$. In principle 
we cannot assure that a normal system does not exist because it may well happen that further derivatives of these 
equations are required to get it. Nonetheless, we proved in proposition \ref{PROPORTIONALITY}
that, if the trace of $P^a_{\ b}$ takes one of the above values, infinite-dimensional Lie algebras of bi-conformal vector 
fields could be constructed from what we conclude that it does not 
exist a normal system for these values of $p$.
 We arrive thus at the following result
\begin{prop}
The only possible cases in which $\Gl(\Su)$ may be infinite dimensional
take place when $P^{a}_{\ b}$ projects on a subspace whose dimension is either $1$, $2$, $n-1$ or $n-2$.
\label{no-closed}
\end{prop}
Note that this completes proposition \ref{PROPORTIONALITY}: infinite dimensional cases can only
arise for square roots generated by 1-forms, 2-forms, $(n-1)$-forms or $(n-2)$-forms. 
\begin{coro}
If $n<6$ then every group of bi-conformal 
transformations is liable to be infinite dimensional. 
\label{six}
\end{coro} 
\section{First integrability conditions: a preliminary analysis}
\label{conditions}
In this section we turn our attention to finding part of the first integrability conditions of the normal system formed by 
(\ref{norm}), its counterpart, and (\ref{normal-set}) together with the constraints (\ref{CONSTRAINT1})-(\ref{CONSTRAINT3}) 
according to the procedure outlined at the beginning of section \ref{highest}. The 
calculation of the full integrability conditions of all the equations 
is long and it has not been completed yet. Nevertheless, we 
have reached a sufficiently advanced stage so that relevant 
information can already be extracted. In this sense, we will  
establish a necessary geometric condition for a line-element to adopt the 
form (\ref{twopieces}) in local coordinates. 
In doing so we will use the calculations performed in the previous section. 

Let us start with the integrability conditions arising from the constraint equations.
The first of such integrability conditions comes from the combination of equations (\ref{18}) with (\ref{9}) yielding
\bea
\fl\lie\left(M_{acb}-\fr{1}{p}E_aP_{bc}+\fr{1}{n-p}W_a\Pi_{cb}\right)=
\phi\left(\Pi_{ap}M^p_{\ cb}-\fr{E_aP_{bc}}{p}\right)+\nonumber\\
\hspace{15mm}+\chi\left(P_{ap}M^p_{\ bc}+\fr{\Pi_{cb}W_a}{n-p}\right).
\label{integral-constraint}
\eea
Observe that (\ref{18}) is the trace of (\ref{9}) so (\ref{integral-constraint}) is the traceless part of (\ref{9}).
This calculation is equivalent to differentiating equation (\ref{CONSTRAINT1}) and using the normal system so 
we can regard (\ref{integral-constraint}) as the first integrability condition of the constraint (\ref{CONSTRAINT1}). 
For a better handling of some forthcoming expressions, let us define the tensor
\be 
T_{abc}\equiv M_{abc}+\fr{1}{n-p}W_a\Pi_{bc}-\fr{1}{p}E_aP_{bc}.
\label{tensor-t}
\ee
Straightforward properties are
$$
P^{ab}T_{abc}=0,\ \Pi^{ab}T_{abc}=0,\ \ P^{bc}T_{abc}=0,\ \ \Pi^{bc}T_{abc}=0,
$$
so $T_{abc}$ is traceless in every index contraction. From this tensor we can define two other as
\bea
A_{abc}\equiv P_{a}^{\ d}T_{dbc}=P_a^{\ d}M_{dcb}+\fr{1}{n-p}W_a\Pi_{cb},\label{splitting-1}\\
B_{abc}\equiv\Pi_{a}^{\ d}T_{dbc}=\Pi_{a}^{\ d}M_{dcb}-\fr{1}{p}E_aP_{cb},\label{splitting-2}
\eea
which allow us to rewrite (\ref{integral-constraint}) in a number of equivalent ways
\bea
\fl\lie T_{abc}=(\phi\Pi_a^{\ s}+\chi P_a^{\ s})T_{sbc}\Longleftrightarrow \lie A_{abc}=\chi A_{abc},\ \lie B_{abc}=\phi B_{abc}
\Longleftrightarrow\label{different-ways}\\
\Longleftrightarrow\lie A^a_{\ bc}=(\chi-\phi)A^a_{\ bc},\ \lie B^a_{\ bc}=(\phi-\chi)B^a_{\ bc},\nonumber
\eea
from what we deduce the invariances
\be
\lie(A^a_{\ bc}B^d_{\ ef})=0,\ \lie(A_{abc}\Pi^{de})=0,\ \lie(B_{abc}P^{de})=0.
\label{spurious-invariance}
\ee
An important property of $T_{abc}$ is shown next.
\begin{theo}
A sufficient condition such that the first integrability condition (\ref{integral-constraint}) of 
(\ref{CONSTRAINT1}) 
is identically satisfied is that the tensor $T_{abc}$ given in (\ref{tensor-t}) vanishes identically. 
\label{GEOMETRIC} 
\end{theo}
\P This follows from the the first of (\ref{different-ways}) straightforwardly.\N

We will see later that the tensor $T_{abc}$ actually vanishes if the metric tensor takes the form (\ref{twopieces}).

The integrability conditions of equations (\ref{CONSTRAINT2}) and 
(\ref{CONSTRAINT3}) present a lengthy form which
shall be omitted in ths paper.
Next we address the integrability conditions of (\ref{norm}-\ref{normal-set}).
 A not very long calculation proves that the integrability conditions 
of the first three equations of (\ref{normal-set})
are identically satisfied. This is evident for the first two
 and is reasonable for the third because the last of (\ref{normal-set}) is, 
in essence, its derivative. 
 Therefore we are only left with the last of (\ref{normal-set}), (\ref{norm}) and its 
counterpart, whose first integrability conditions are calculated by means of the Ricci identity 
(\ref{ricci}) applied to $\nb_{[a}\nb_{b]}\Psi_{cd}$, $\nb_{[a}\nb_{b]}\phi_c$ and $\nb_{[a}\nb_{b]}\chi_c$, 
respectively, and using the system itself to get rid of the first derivatives
of the system variables. This last calculation and the 
geometric conditions imposed by the whole set of first integrability conditions are
 under current research and they will be presented elsewhere.
 
\subsection{The condition $T_{abc}=0$}
The vanishing of the tensor $T_{abc}$ is a sufficient condition for the integrability constraint 
(\ref{integral-constraint})
to be fulfilled as we proved in theorem \ref{GEOMETRIC}. To investigate further the geometric significance of this 
condition, let us compute this tensor for metrics given in local coordinates $\{x^{a}\}$ by
\be
\fl ds^2=\rmg_{ab}dx^{a}dx^{b}=\rmg_{\alpha\beta}dx^{\alpha}dx^{\beta}+\rmg_{AB}dx^Adx^B,\ \ 
\alpha,\beta=1,\dots p,\ \ A,B=p+1,\dots, n,
\hspace{.2cm}
\label{particion}
\ee
where $\rmg_{\alpha\beta}$ and $\rmg_{AB}$ are functions of all the coordinates $\{x^{a}\}$.
The tensors $P_{ab}$ and $\Pi_{ab}$ with components
\be
P_{ab}=\rmg_{\alpha\beta}\delta^{\alpha}_{\ a}\delta^{\beta}_{\ b},\ \ \ 
\Pi_{ab}=\rmg_{AB}\delta^A_{\ a}\delta^B_{\ b},
\ee    
are orthogonal projectors, whose nonvanishing components are 
$P_{\alpha\beta}=\rmg_{\alpha\beta}$ and $\Pi_{AB}=\rmg_{AB}$, playing the role of (\ref{losproyectores}) 
in this case. Thus, they can be used to calculate $T_{abc}$ according 
to its definition (\ref{tensor-t}). The non-zero components of the Christoffel 
symbols are
\bnr
\fl\Chr^{\alpha}_{\ \beta\g}=
\fr{1}{2}\rmg^{\alpha\rho}(\d_{\beta}\rmg_{\g\rho}+\d_{\g}\rmg_{\rho\beta}-\d_{\rho}\rmg_{\beta\g}),\ 
\Chr^{\alpha}_{\ \beta A}=\fr{1}{2}\rmg^{\alpha\rho}\d_A\rmg_{\beta\rho},
\ \Chr^\alpha_{\ BA}=-\fr{1}{2}\rmg^{\alpha\rho}\d_{\rho}\rmg_{BA},\\
\fl\Chr^A_{\ B\alpha}=\fr{1}{2}\rmg^{AD}\d_{\alpha}\rmg_{BD},\ 
\Chr^A_{\ \alpha\beta}=-\fr{1}{2}\rmg^{AD}\d_D\rmg_{\beta\alpha},\ 
\Chr^A_{\ BC}=\fr{1}{2}\rmg^{AD}(\d_B\rmg_{CD}+\d_C\rmg_{DB}-\d_D\rmg_{BC}),
\enr
from what we conclude that the nonvanishing components of $M_{abc}$, 
$E_a$, and $W_a$ are, respectively
\bnr
M_{\alpha AB}=\d_{\alpha}\rmg_{AB},\ \ 
M_{A\alpha\beta}=-\d_{A}\rmg_{\alpha\beta}, \\
E_{A}=-\d_{A}\log|\mbox{det}(\rmg_{\alpha\beta})|,\hspace{3mm}
W_{\alpha}=-\d_{\alpha}\log|\mbox{det}(\rmg_{AB})|.
\enr
The condition $T_{abc}=0$ entails the couple of partial differential equations
$$
\d_{\alpha}\rmg_{AB}=-\, \fr{1}{n-p}\, \rmg_{AB}W_{\alpha},\ \ 
\d_A\rmg_{\alpha\beta}=-\, \fr{1}{p}\, \rmg_{\alpha\beta}E_A,
$$
whose solution is 
\be
\rmg_{\alpha\beta}=G_{\alpha\beta}(x^{\delta})e^{\Lambda_1(x^a)},\ 
\rmg_{AB}=G_{AB}(x^D)e^{\Lambda_2(x^a)},
\label{splitting}
\ee
where $G_{\alpha\beta}$, $G_{AB}$, $\Lambda_1$, $\Lambda_2$ 
are arbitrary functions of their respective arguments 
 with no restrictions other than det$(G_{\alpha\beta})\neq 0$, det$(G_{AB})\neq 0$. 
We arrive thus at the following important result.
\begin{theo}
A necessary condition for the existence of a coordinate system in which
a metric $\rmg_{ab}$ decomposes according to equation (\ref{particion}) with $\rmg_{\alpha\beta}$ and 
$\rmg_{AB}$ given by (\ref{splitting}) is that
the tensor $T_{abc}$ defined in (\ref{tensor-t}) vanishes identically.
\label{decomposition1}\N
\end{theo}
{\bf Remark}. Observe once again that this theorem holds even in the 
cases with $p$ and/or $n-p$ taking the values 1 or 2.

The vanishing of the tensor $T_{abc}$ is thus part of a possible invariant characterization 
of ``breakable spaces'', in the sense that the metric tensor decomposes 
according to (\ref{particion}) and (\ref{splitting}). These 
spaces have been called {\em double-twisted products} in 
\cite{PONGE}, as the ``warping'' or ``twisting'' functions $\Lambda_1$ and $\Lambda_2$ depend on all  
coordinates of the manifold. Particular
interesting cases of these are (i) warped-product spaces where $\Lambda_1,\Lambda_2$ depend only on the 
$\{x^{\alpha}\}$ coordinates, say,
(see e.g. \cite{BO}, or \cite{BEEM} for a study of warped product spaces in Lorentzian geometry);
(ii) conformally reducible spaces in which   
$e^{\Lambda_1}=e^{\Lambda_2}$ (see \cite{CAROT-TUPPER} for dimension four and Lorentzian signature),
(iii) twisted-product spaces where $\Lambda_1$ (say) depends only on the coordinates 
$x^{\alpha}$ ---so that $g_{\alpha\beta}$ is a ``true'' metric on the 
$\{x^{\alpha}\}$-space---, see \cite{CHEN,PONGE,KUPELI}; and (iv)
double-warped product spaces, e.g. \cite{PONGE,CAROT2}, where $\Lambda_1$ 
depends only on the $\{x^A\}$-coordinates and vice versa for $\Lambda_2$. 
A major problem when dealing with such breakable spaces is their invariant characterization by means of 
some (local) criterion. Attempts 
towards this direction have been made for instance in
\cite{CAROT,CAROT-TUPPER,CAROT2}. 

In the particular case of  $G_{\alpha\beta}$ and $G_{AB}$ being conformally flat metrics we recover the case studied in 
proposition \ref{canonical-form} and the space is maximal, namely, it
admits a maximum number of bi-conformal vector fields.

\section{Examples}
\label{examples}
{\bf Example 1.} Our first example of bi-conformal vector field was already presented in \cite{PIII}. 
We briefly mention it here due to its interest in Lorentzian 
geometry (see \cite{PIII} for the definitions of the concepts and the proof).
\begin{prop}
Every causal-preserving vector field with $n-1$ linearly independent canonical null directions
for $n\geq 3$ is a bi-conformal vector field with $\Su$ the Lorentz 
tensor built up from the canonical null directions.
\label{n-1}
\end{prop}

\medskip
\noindent
{\bf Example 2.} We present next an example of an algebra of bi-conformal 
vector fields which generalizes the one we gave in \cite{LETTER} for warped product spacetimes. 
A bi-conformal vector field is fixed once we set a simple $p$-form 
$\O=\z_1\wedge\dots\wedge\z_p$ defining a square root $\Su$. A simple choice
 is a differential $p$-form such that the distribution spanned by 
$\z_1,\dots,\z_p$ is integrable which, according to Frobenius theorem, happens 
if and only if $d\z_{\alpha}=\sum_{\beta=1}^p\bomega_{\alpha\beta}\wedge\z_{\beta}$ for a certain set of 1-forms 
$\bomega_{\alpha\beta}$. This means that $d\O=\bomega\wedge\O$, 
$\bomega=\bomega_{11}-\bomega_{22}+\dots+(-1)^{p-1}\bomega_{pp}$ 
and we can set up a coordinate system $\{x^1,\dots, x^{n}\}$ in which $\O$ takes the form 
$\O=\rho(x)dx^1\wedge\dots\wedge dx^{p}$. The line element 
in this coordinate system can be decomposed as
$$
ds^2=\rmg_{\alpha\beta}dx^{\alpha}dx^{\beta}+2\rmg_{\alpha A}dx^{\alpha}dx^A
+\rmg_{AB}dx^Adx^B,\ \ 
1\leq \alpha,\beta\leq p,\ 
p+1\leq A,B\leq n,
$$
where det($\rmg_{\alpha\beta})\neq 0$ and the signatures of $\rmg_{\alpha\beta}$ and $\rmg_{AB}$ are left free. 
The metric tensor components depend on all coordinates. Formula (\ref{superenergy}) provides $S_{ab}$ once 
we know $\O$ yielding
\be
S_{ab}dx^{a}dx^{b}=\sigma_{\alpha\beta}dx^{\alpha}dx^{\beta}-
2\rmg_{\alpha A}dx^{\alpha}dx^A-\rmg_{AB}dx^Adx^B,
\ee
where $\sigma_{\alpha\beta}$ can be determined in terms of $\rmg_{\alpha\beta}$, $\rmg_{AB}$ 
and $\rmg_{\alpha A}$ using (\ref{superenergy}). Now, as we proved in proposition 
\ref{second-form}, the second equation of (\ref{fundamental}) is equivalent to $\lie\O=p(\alpha+\beta)\O/2$ 
and this last equation takes the form
\bnr
\fl\lie(\rho dx^1\wedge\dots\wedge dx^{p})=(\lie\rho)dx^1\wedge\dots\wedge dx^{p}+
\rho d\xi^1\wedge\dots\wedge dx^{p}+\dots+\\
\fl+\rho dx^1\wedge\dots\wedge d\xi^{p}
=(\lie\rho+\d_{\alpha}\xi^{\alpha})dx^1\wedge\dots\wedge dx^{p}+
\rho\d_A\xi^{1}dx^A\wedge dx^2\wedge\dots\wedge dx^{p}+\\
\fl+\dots+\rho\d_A\xi^{p}dx^1\wedge\dots\wedge dx^{p-1}\wedge dx^A=
\fr{p}{2}(\alpha+\beta)\rho dx^1\wedge\dots\wedge dx^{p},
\enr  
from what we conclude
\bea
\fr{1}{\rho}\lie\rho+\d_{\alpha}\xi^{\alpha}=\fr{p}{2}(\alpha+\beta),\label{derivadadero}\\ 
\d_B\xi^{\alpha}=0\ \ \Rightarrow\xi^{\alpha}=\xi^{\alpha}(x^{\beta}).
\label{p-m-eq}
\eea
This is our first set of equations. The remaining ones come from the first equation of 
(\ref{fundamental}) which written in components looks like
\be
\xi^{a}\d_{a}\rmg_{mn}+\d_{m}\xi^{r}\rmg_{rn}+
\d_{n}\xi^{r}\rmg_{mr}=a\rmg_{mn}+\beta S_{mn}.
\label{munu}
\ee
If we spell out the different components, we get
\bea
\fl\xi^{a}\d_{a}\rmg_{\alpha\beta}+\d_{\alpha}\xi^{\g}\rmg_{\g\beta}+
\d_{\beta}\xi^{\g}\rmg_{\alpha\g}+\d_{\alpha}\xi^B\rmg_{B\beta}+
\d_{\beta}\xi^B\rmg_{\alpha B}=\alpha\rmg_{\alpha\beta}+
\beta\sigma_{\alpha\beta}\\
\fl\xi^{\alpha}\d_{\alpha}\rmg_{AB}+\xi^C\d_{C}\rmg_{AB}+\d_A\xi^C\rmg_{CB}+
\d_{B}\xi^C\rmg_{AC}=(\alpha-\beta)\rmg_{AB}\\
\fl\xi^{\g}\d_{\g}\rmg_{A\alpha}+\xi^B\d_B\rmg_{A\alpha}+
\d_A\xi^B\rmg_{B\alpha}+\d_{\alpha}\xi^{\g}\rmg_{A\g}
+\d_{\alpha}\xi^B\rmg_{AB}=(\alpha-\beta)\rmg_{\alpha A}.\label{prepar3}
\eea
It is convenient to split the vector field $\xiv$ in two parts 
$$
\xiv_1=\xi^{\alpha}\d_{\alpha},\ \  
\xiv_2=\xi^A\d_A,
$$ 
so that $\xiv=\xiv_1+\xiv_2$. Observe that $\xiv_1$ is a genuine vector field in the distribution spanned 
by $\O$.
 In terms of them the previous equations are rewritten as
\bea
(\liea\rmg)_{\alpha\beta}+(\lieb\rmg)_{\alpha\beta}=
\alpha\rmg_{\alpha\beta}+\beta\sigma_{\alpha\beta}\label{par1}\\
(\liea\rmg)_{AB}+(\lieb\rmg)_{AB}=(\alpha-\beta)\rmg_{AB}\label{par2}\\
(\liea\rmg)_{A\alpha}+(\lieb\rmg)_{A\alpha}=(\alpha-\beta)\rmg_{A\alpha},
\label{par3}
\eea
These are the most general set of equations for 
bi-conformal vector fields such that 
the projector $P_{ab}$ can be regarded as conformally related to a true metric tensor 
defined on a submanifold of the total manifold $(V,\G)$. The only nonvanishing components of this projector are
$$
P_{\alpha\beta}=\fr{1}{2}(\s_{\alpha\beta}+\rmg_{\alpha\beta}).
$$
An interesting case arises when $\rmg_{A\alpha}=0$ or in other words the 
metric tensor $\rmg_{ab}$ is breakable 
in two parts as in (\ref{particion}), each of them depending on all the coordinates. 
 From equation (\ref{par3}) (see (\ref{prepar3})) we get that 
$$
\rmg_{A\alpha}=0\Rightarrow\d_{\alpha}\xi^{B}=0
$$
so that $\xiv_2$ depends only on the coordinates $\{x^{A}\}$ and it is a genuine vector field of the distribution spanned by
$\{\d/\d x^{A}\}$. Furthermore $\s_{\alpha\beta}$ turns out to be $\rmg_{\alpha\beta}$ so (\ref{par1}) and (\ref{par2}) become
\bea
\liea\rmg_{\alpha\beta}+\lieb\rmg_{\alpha\beta}=(\alpha+\beta)\rmg_{\alpha\beta}\\
\lieb\rmg_{AB}+\liea\rmg_{AB}=(\alpha-\beta)\rmg_{AB},
\eea
where the only nonvanishing components of the projectors $P_{ab}$ and $\Pi_{ab}$ are
 $P_{\alpha\beta}=\rmg_{\alpha\beta}$ and $\Pi_{AB}=\rmg_{AB}$. This couple of equations can be solved in an 
adapted coordinate system where $\xiv_1$ and $\xiv_2$ take the form $\xiv_1=\d/\d x^1$, $\xiv_2=\d/\d x^n$. The 
solution consists on a metric tensor such that the two pieces $\rmg_{\alpha\beta}$ and $\rmg_{AB}$  can be factored as 
\be
\rmg_{\alpha\beta}=f\, G_{\alpha\beta},\ \ \ \rmg_{AB}=h\, G_{AB},
\label{descomposiciontotal}
\ee
where $0=\lie G_{\alpha\beta}=\lie G_{AB}$ and $f$, $h$ are nonvanishing 
functions otherwise arbitrary.
This can be compared with the form found in proposition \ref{adapted} for a general metric 
in coordinates adapted to $\xiv$. 
 
Equation (\ref{descomposiciontotal})
is more general than (\ref{splitting}) as $G_{\alpha\beta}$ and $G_{AB}$ 
may depend on all coordinates: this is just a breakable space, 
not necessarily double-twisted. Nonetheless, the considerations pointed out after 
theorem \ref{decomposition1} hold also here.
Interesting subcases  of (\ref{descomposiciontotal}) using warped-product 
spacetimes were treated 
at the end of \cite{LETTER}.

\medskip
\noindent
{\bf Example 3.} In Lorentzian geometry, let us consider a bi-conformal vector field such that 
$S_{ab}$ is the Lorentz tensor of a normalized timelike 1-form $u_a$. In physical terms this 1-form 
may represent the velocity vector field of a fluid, or the congruence associated to a set of observers, 
or a reference system, among others.
In this case the explicit form of $S_{ab}$ is found to be 
$S_{ab}=-\rmg_{ab}+2u_au_b$ and equations (\ref{bi-conformal}) (or (\ref{semi-conformal}))  become
\be
\fl\lie(u_au_b)=(\alpha+\beta)u_au_b, \ \ \ \ 
\lie(\rmg_{ab}-u_au_b)=(\alpha-\beta)(\rmg_{ab}-u_au_b),
\label{general-selfsimilar}
\ee
from where
\be
\lie u_a=\fr{1}{2}(\alpha+\beta)u_a.
\label{SELF-SIMILAR}
\ee 
The tensor $h_{ab}\equiv\rmg_{ab}-u_au_b$ is the orthogonal projector defined by 
the congruence whose tangent vector is $u^a$. It is worth remarking here that the 
case with $\alpha$ and $\beta$ fixed constants is a symmetry already 
known in the literature as {\it kinematic self-similarity}. It was first introduced in 
\cite{CARTER} and was later studied  
specially in spherically-symmetric perfect-fluid spacetimes 
\cite{SINTES,BENOIT,CCOLEY}. Kinematic self-similarity can
be interpreted as saying that the flow generated by $\xiv$ scales by unequal 
factors the timelike direction $u^a$ and the spacelike directions contained in
$h_{ab}$ as opposed to {\it self-similarity} where $\xiv$ is a homothetic Killing vector and 
the scaling takes place by the same factor in all the directions. 
As we see, (\ref{general-selfsimilar}) and (\ref{SELF-SIMILAR}) are a generalization of
kinematic self-similarity --- which is obviously included --- such that the gauges are non-constant, 
so that the solutions for $\xiv$ could be called ``kinematic conformal vector''.
 
 Another interesting case of the above equations arises when $\xi^{a}=u^{a}$. 
In this case the acceleration 1-form $a_a$ of the congruence defined by the vector field $u^a$ is given by 
$\lieu u_a=a_a$ from what we deduce, using (\ref{SELF-SIMILAR}) and 
the orthogonality $a_{a}u^a=0$, that $a_a=0$ i.e. this is a geodesic 
congruence. Moreover, 
$\alpha=-\beta$ and the second equation of (\ref{general-selfsimilar}) becomes 
$$
\lieu h_{ab}=2\alpha h_{ab}
$$
which means that this is a shear-free congruence whose expansion is proportional 
to $\alpha$ (see e.g. \cite{KRAMERS} for definitions). 
In other words, every geodesic shear-free timelike
congruence defines a bi-conformal vector field where the Lorentz tensor is $\Su=\T\{\vu\}$. These congruences have been 
extensively studied in General Relativity specially for perfect fluids whose velocity vector is $u^a$  
(see e.g. \cite{COLLINS,SZEKERES,KRASINSKI,JOURNALOFPHYSICSC}). 
It is known \cite{SZEKERES} that every such {\em perfect-fluid} congruence 
is either expansion-free or irrotational. In the former case, $\alpha=0$ necessarily and 
the congruence is a {\em geodesic rigid motion} having the remarkable geometric property that 
it defines a {\em homogeneous} family of observers i.e. the average distance between neighbouring observers remains 
the same along the congruence. These congruences
 were first defined by Born \cite{BORN} as the relativistic generalization of the rigid motions used in Newtonian mechanics
 and further studies of them can be found in \cite{BELLOSA,BELLOSA2,BEL-ERE,MASON,PIRANI,WAHLQUIST,SYNGE} 
and references therein. The latter case can only be achieved if the perfect 
fluid defines a Robertson-Walker solution and again the 
fluid vector congruence defines a privileged observer.   

\section*{Acknowledgements}
We thank comments from M. S\'anchez. 
Financial support of the Basque Country University and the Spanish Ministry of 
Science and Technology under the grants 9/UPV00172.310-14456/2002 and 
BFM 2000-0018 is acknowledged. 
\appendix

\section*{Appendix A: Number of independent constraints in equation (\ref{CONSTRAINT1})}
\label{APPENDIX}
\setcounter{section}{1}
\setcounter{equation}{0}
In this appendix we will show how many linearly independent equations
are contained in (\ref{CONSTRAINT1}). This equation  
 can be cast in the following equivalent form
\be
\xi_c\nb^cS_{ab}+M_{ab}^{\ \ qc}\Psi_{qc}=0,\ \ 
M_{ab}^{\ \ qc}=\delta^{[q}_{\ a}S^{c]}_{\ b}+\delta^{[q}_{\ b}S^{c]}_{\ a}=
2\delta^{[q}_{\ (a}S^{c]}_{\ b)}
\label{system}
\ee               
We can regard this equation as an homogeneous system for the variables $\xi_c$ and
$\Psi_{qc}$ where the indexes $qc$ and $ab$ are gathered in a block and thus
considered as a single index as far as the calculations are concerned. Therefore
we can rewrite (\ref{system}) in an explicit matrix form
$$
\left(\begin{array}{cc}
         (\nb\Su)_{I}^{\ c} & M_{I}^{\ K} \end{array}\right)
\left(\begin{array}{c}
                 \xi_{c}\\                        
                \Psi_{K}
                   \end{array}\right)=0,\ \  I=\{ab\},\ K=\{qc\},
$$ 
where $1\leq K\leq n(n-1)/2$, $1\leq I\leq n(n+1)/2$, $1\leq c\leq n$. 
Hence the number of linearly independent equations present in (\ref{system})
is given by the rank of the matrix of this homogeneous system. In order to calculate 
the rank of this matrix, we choose an 
orthonormal basis $\{\vec{\e}_1,\dots,\vec{\e}_{n}\}$ of eigenvectors of $S^a_{\ b}$ and label them according
to the following scheme
$$
\G(\vec{\e}_{\alpha},\vec{\e}_{\alpha})=+1,\ \G(\vec{\e}_{A},\vec{\e}_{A})=-1,\ \ 1\leq\alpha\leq r,\ r+1\leq A\leq n.
$$  
Thus in this basis $\G=\mbox{diag}(\overbrace{1,\dots,1}^{r},\overbrace{-1,\dots,-1}^{n-r})$. In the remaining parts of 
this appendix we will follow the same convention used to label the elements of the orthonormal basis, namely, Greek indexes
will run from 1 to $r$ and capital Latin indexes from $r+1$ to $n$.  
 In the above orthonormal basis the tensor $S^a_{\ b}$ looks like (see proposition \ref{spectral}) 
$$
S^a_{\ b}=\left(\begin{array}{cc}
                              \mathbb{I}_{p\times p} &\ \\
                               \         & -\mathbb{I}_{(n-p)\times (n-p)}
                    \end{array}\right),
$$    
where $\mathbb{I}_{m\times m}$ is the $m$-dimensional identity matrix. It is
then clear that in this basis the only non vanishing components of 
$M_{ab}^{\ \ qc}$ are those with $a=q$, $b=c$ being the value of such components
$$
M_{ab}^{\ \ ab}=\fr{1}{2}(\delta^a_{\ a}S^b_{\ b}-\delta^b_{\ b}S^a_{\ a})=
S^b_{\ b}-S^a_{\ a},\ a< b\ \ \mbox{(no summation).}
$$ 
All these components have different row and column indexes so the rank of $M_{I}^{\ K}$
is the total number of such components. Given
the form of $S^a_{\ b}$ we find by a simple counting that such number 
turns out to be $p(n-p)$.

Let us now spell out the covariant
derivative of $S_{ab}$ in the above chosen orthonormal basis
$$
\nb_cS_{ab}=-\g^e_{ca}S_{eb}-\g^e_{cb}S_{ae}
=-\g^b_{ca}S_{bb}-\g^a_{cb}S_{aa},
$$
where $\g^a_{bc}$ are the components of the connection in this basis
and the last step follows using that $S_{ab}$ is a diagonal tensor in 
this basis (there is no summation in the last step). In the above orthonormal frame, 
the connection components satisfy the symmetry properties
$$
\g^a_{ba}=0,\ \g^{\alpha}_{bB}=\g^B_{b\alpha},\ \g^{\alpha}_{c\beta}+\g^{\beta}_{c\alpha}=0,\ \g^{A}_{bB}+\g^B_{bA}=0, 
$$
from what we conclude that the only nonvanishing components of $\nb_cS_{ab}$ are 
(again there is no summation in the repeated 
indexes)
$$
\nb_cS_{B\alpha}=-\g^{\alpha}_{cB}(S_{\alpha\alpha}+S_{BB}),\ 
\nb_cS_{AB}=-\g^B_{cA}(S_{BB}-S_{AA}),\ 
\nb_cS_{\alpha\beta}=-\g^{\beta}_{c\alpha}(S_{\beta\beta}-S_{\alpha\alpha}). 
$$
 Comparing these equations with the above formulae for $M_{I}^{\ K}$ we infer that 
$M_I^{\ K}=0$ $\forall K$ implies that $(\nb\Su)_I^{\ c}$ vanishes for every
value of the index $c$ so the rank of the matrix system is just the rank of 
$M_I^{\ K}$ calculated above.


\section*{References}
\begin{thebibliography}{99}
\bibitem{BEEM} Beem J K, Ehrlich P E, and Easley K L 1996 {\it Global Lorentzian Geometry}
 (Pure and Applied Math. 202, Marcel Dekker, New York)
\bibitem{BEL-ERE} Bel L 1990, in {\em Recent Developments in Gravitation}, 
E. Verdaguer, J. Garriga and J. C\'espedes eds. (World Scientific, Singapore) 
\bibitem{BELLOSA} Bel L and Llosa J 1995 {\it Class. Quantum Grav.} {\bf 12} 1949
\bibitem{BELLOSA2} Bel L and Llosa J 1995 {\em Gen. Rel. Grav.} {\bf 27} 1089 
\bibitem{BENOIT} Benoit P M and Coley A A 1998 {\it Class. Quantum Grav.} {\bf 15} 2397 
\bibitem{PI} Bergqvist G and Senovilla J M M  2001 {\it Class. Quantum Grav.}
{\bf 18} 5299
\bibitem{BILYALOV} Bilyalov R F 1964 {\em Sov. Phys.} {\bf 8} 878
\bibitem{BO} Bishop R L and O'Neill B 1969 {\it Trans. Amer. Math. Soc.} {\bf 145} 1
\bibitem{BORN} Born M 1909 Phys. Z. {\bf 10} 814
\bibitem{CAROT} Carot J and da Costa J 1993 {\it Class. Quantum Grav.} {\bf 10} 461
\bibitem{CAROT-TUPPER} Carot J and Tupper B O J 2002 {\it Class. Quantum Grav.} {\bf 19} 4141
\bibitem{CARTER} Carter B and Henriksen R N 1989 {\em Class. Quantum Grav.} {\bf 15} 2397 
\bibitem{CHEN} Chen Bang-yen 1981 {\it Geometry of submanifolds and its applications} (Science Univ. Tokyo, Tokyo)
\bibitem{CHERN} Chern S S Chen W H and Lam K S 1999 {\it Lectures on Differential Geometry}
(World Scientific Publishing)   
\bibitem{CCOLEY} Coley A A 1997 {\it Class. Quantum Grav.} {\bf 14} 87 
\bibitem{KERR-SCHILD} Coll B, Hildebrandt S R and Senovilla J M M 2001
{\it Gen. Rel. Grav.} {\bf 33} 649
\bibitem{JOURNALOFPHYSICSC} Collins C B 1986 {\em Can. J. Phys.} {\bf 64} 191 
\bibitem{COLLINS} Collins C B and Lang J M 1987 {\it Class. Quantum Grav} {\bf 4} 61
\bibitem{DF} Defrise-Carter L 1975 {\it Commun. Math. Phys.} {\bf 40} 273 
\bibitem{EISENHARTII} Eisenhart L P 1933 {\it Continuous groups of transformations} 
(Dover publications Inc. New York)  
\bibitem{KUPELI} Fern\'andez-L\'opez M, Garc\'{\i}a-R\'{\i}o E and Kupeli D N 2001 {\em Manuscripta Math.} 
{\bf 106} 213
\bibitem{CAUSAL} Garc\'{\i}a-Parrado A and Senovilla J M M 2003 {\it Class. Quantum Grav.} {\bf 20} 625
\bibitem{LETTER} Garc\'{\i}a-Parrado A and Senovilla J M M 2003 {\it Class. Quantum Grav.} {\bf 20} L139 
\bibitem{PIII} Garc\'{\i}a-Parrado A and Senovilla J M M 2004 {\it Class. Quantum Grav.} {\bf 21} 661 
\bibitem{HALL1} Hall G S 1988, in {\it Relativity Today}, Z. Perjes ed. (World Scientific, Singapore)
\bibitem{HALL-DEFRISE1} Hall G S 1990 {\em J. Math. Phys.} {\bf 31} 1198
\bibitem{HALL2} Hall G S 1998 {\em Gen. Rel. Grav.} {\bf 30} 1099
\bibitem{HALL-DEFRISE2} Hall G S and Steele J D 1991 {\em J. Math. Phys.} {\bf 32} 1847
\bibitem{sergi} Hildebrandt S R 2002 {\it Gen. Rel. Grav.} {\bf 34} 65
\bibitem{LEVINE} Katzin G H, Levine J and Davis W R 1969 {\it J. Math. Phys.} {\bf 10} 617
\bibitem{KRASINSKI} Krasi\'nski A  1996 {\em Inhomogeneous cosmological models}
 (Cambridge University Press, Cambridge)
\bibitem{MASON} Mason D P and Pooe C A 1987 {\em J. Math. Phys.} {\bf 28} 2705
\bibitem{PIRANI} Pirani F and Williams G 1961-62 Rigid motion in a gravitational field 
{\em Seminaire Janet (M\'ecanique analytique et M\'ecanique c\'eleste)}
{\bf 8-9}
\bibitem{PONGE} Ponge R and Reckziegel H 1993 {\em Geom. Dedicata} {\bf 48} 15
\bibitem{CAROT2} Ramos M, Vaz E and Carot J 2003 {\em J. Math. Phys.} {\bf 44} 4839
\bibitem{SUP} Senovilla J M M 2000 {\it Class. Quantum Grav.} {\bf 17} 2799
\bibitem{SZEKERES} Senovilla J M M, Sopuerta C F and Szekeres P 1998 {\em Gen. Rel. Grav.} {\bf 30} 389
\bibitem{SINTES} Sintes A M 1998 {\it Class. Quantum Grav.} {\bf 15} 3689
\bibitem{KRAMERS} Stephani H, Kramer D, MacCallum M A H, Hoenselaers C and Herlt E 2003 
{\it Exact Solutions to Einstein's Field Equations Second Edition}
(Cambridge University Press, Cambridge)
\bibitem{SYNGE} Synge J L 1965 {\em Relativity: the special theory} (North Holland, Amsterdam)
\bibitem{TSAMPARLIS} Tsamparlis M 1992 {\it J. Math. Phys.} {\bf 33} 1472 
\bibitem{WAHLQUIST}  Wahlquist H D and Eastbrook F B 1966 {\em J. Math. Phys.} {\bf 7} 894 
\bibitem{YANO} Yano K 1955 {\it Theory of Lie Derivatives} (North Holland) 
\bibitem{ZAFIRIS} Zafiris E 1997 {\it J. Math. Phys.} {\bf 38} 5854 
\end{thebibliography}
\end{document}